# Analytical solution to swing equations in power grids

**HyungSeon Oh** *

Department of Electrical and Computer Engineering, United States Naval Academy, Annapolis, Maryland, United States of America

* hoh@usna.edu

**Data Availability Statement:** Illinois Center for a Smarter Electric Grid (ICSEG) Available at https://icseg.iti.illinois.edu/power-cases/

**Funding:** The author received no specific funding for this work.

**Competing interests:** The author has declared that no competing interests exist.

## Abstract

### Objective

To derive a closed-form analytical solution to the swing equation describing the power system dynamics, which is a nonlinear second order differential equation.

### Existing challenges

No analytical solution to the swing equation has been identified, due to the complex nature of power systems. Two major approaches are pursued for stability assessments on systems: (1) computationally simple models based on physically unacceptable assumptions, and (2) digital simulations with high computational costs.

### Motivation

The motion of the rotor angle that the swing equation describes is a vector function. Often, a simple form of the physical laws is revealed by coordinate transformation.

### Methods

The study included the formulation of the swing equation in the Cartesian coordinate system, which is different from conventional approaches that describe the equation in the polar coordinate system. Based on the properties and operational conditions of electric power grids referred to in the literature, we identified the swing equation in the Cartesian coordinate system and derived an analytical solution within a validity region.

### Results

The estimated results from the analytical solution derived in this study agree with the results using conventional methods, which indicates the derived analytical solution is correct.

### Conclusion

An analytical solution to the swing equation is derived without unphysical assumptions, and the closed-form solution correctly estimates the dynamics after a fault occurs.

## Nomenclature

| | |
|---|---|
| CCS | Cartesian coordinate system |
| COI | center of inertia |
| COM | coupled oscillation model |
| DAE | differential algebraic equation |
| DM | direct method |
| IEEE | Institute of Electrical and Electronics Engineers |
| MVA | Mega Volt Amp |
| PCS | polar coordinate system |
| TDS | time domain simulation |
| $\gamma_{ik}$ | admittance angle of the line connecting Buses *i* and *k* |
| $\gamma^K$ | cohesive angle for the coupled oscillation model |
| $\kappa_F()$ | condition number of the matrix inside the parenthesis associated with the Frobenius norm, $\kappa_F(M)=\lVert M\rVert_F \lVert M^{-1}\rVert_F$ |
| $\delta_i$ | rotor angle of a generator *i* |
| λ | eigenvalue |
| $\vartheta()$ | order of the quantity inside the parenthesis |
| $\varsigma^K$ | coupling strength in the Kuramoto model |
| $\theta_k$ | terminal voltage angle at Bus *k* that a generator *i* is directly connected to |
| $\delta_i^K$ | phase of the $i^{\text{th}}$ oscillator in the Kuramoto model |
| $\omega_i$ | $d\delta_i/dt$ the speed of rotor angle of a generator *i* |
| $\omega_i^K$ | natural frequency of the $i^{\text{th}}$ oscillator in the Kuramoto model |
| $D_i$ | damping term associated with a generator *i* |
| $E_i$ | voltage magnitude of the $i^{\text{th}}$ *Ibus* that remains unchanged over the transient |
| $E_k$ | terminal voltage magnitude |
| *Ibus* | bus representing a generator |
| *Kbus* | bus that is directly connected to a generator or generators, i.e., slack bus or *PV* bus |
| *Mbus* | bus that is not directly connected to a generator, i.e., *PQ* bus |
| $M_{ref}$ | reference inertia assigned for a frequency- or time-dependent load |
| $M_i$ | inertia of a rotor in a generator *i* |
| *Nb* | number of terminal buses in the system including *Kbus* and *Mbus* |
| *ND* | number of frequency- or time-dependent loads |
| *NI* | number of generators in the system including *Ibus* |
| *Y* | admittance matrix |
| bldiag | block-diagonal matrix |
| diag | diagonal matrix |
| $e_i$ | $i^{\text{th}}$ column vector in an identity matrix |
| $g_{ii}$ | real component of the line admittance connecting a generator *i* and Bus *k* |
| *h* | scaling factor to adjust the update of $d^2w_I/dt^2$ |
| $i_j$ | injection current at Bus *j* |
| $i_j^0$ | injection currents at Bus *j* when no voltages are applied |
| $p_i$ | real power injection from a generator *i* |
| $p_{i\to k}^{elec}$ | electrical power output from a generator *i* to Bus *k* |
| $p_i^{mech}$ | mechanical power input to a generator *i* |
| $p_{max}$ | maximum power injection, $p_{\max} = \lvert \tilde{y}_{ik} \rvert E_i E_k$ |
| $q_i$ | reactive power injection from a generator *I* to Bus *k* |
| rad | radian |
| $s(\lambda)$ | condition of λ, $s(\lambda) = \lvert u_L^H u_R \rvert$ |
| $t_0$ | time when a disturbance (a physical anomaly and/or a control action) occurs |
| $u_L$ | left eigenvector of *Mat*, $u_L^H Mat = \lambda u_L^H$ |
| $u_R$ | right eigenvector of *Mat*, $Mat\, u_R = \lambda u_R$ |
| $v_j$ | voltages at Bus *j* |
| $v_x$ | real part of the voltages *v* |
| $v_y$ | imaginary part of the voltages *v* |
| $x_I$ | real part of the loss-reflected voltage at *Ibuses* |
| $y_I$ | imaginary part of the loss-reflected voltage at *Ibuses* |
| $\lvert y_{ik} \rvert$ | magnitude of line admittance connecting a generator *i* and Bus *k* |

# Introduction

Electric power loads are expected to be fulfilled continuously in modern society, and when a load is not satisfied it is termed an "event". The infrastructure to generate and transport electricity to end-consumers is called a power system. In the United States, this infrastructure comprises 19,023 individual, commercial generators (6,997 power plants) [1], 70,000 substations [2], and 360,000 miles of lines [3]. The number of power electronic devicess is more than a million, and non-anticipated losses of system components inevitably occur. The Federal Energy Regulatory Commission in the United States regulates the interstate transmission and wholesale of electricity. According to a report submitted to them [4], the 1-in-10 standard is a widely used reliability standard across North America. To meet this standard in a large-scale network with many system components, the power system must be able to withstand sudden disturbances (such as electric short circuits or non-anticipated loss of system components). It should be noted that most disturbances (including the failure of components) do not lead to an event. When a disturbance occurs, a governor regulates the speed of a machine to adjust the output power of a generator according to the network conditions. In general, the timeframe of governor action is approximately 0.1–10 s [5]. Therefore, it is necessary to assess if the power system is stable approximately 10 s after a disturbance occurs, which is the subject of the transient stability assessment.

Viewed overall, power systems consist of mechanical and electrical systems that obey energy conservation and Kirchhoff's laws, which are integrated as the so-called swing equation [5],

$$M_i \frac{d^2\delta_i}{dt^2} + D_i \frac{d\delta_i}{dt} = p_i^{mech} - p_{i\to k}^{elec} .$$

The swing equation is a heterogeneous nonlinear second-order differential equation with multi-variables. There is no known method to solve the differential equation in an analytical fashion. Instead, several approaches to analyze the problem are suggested: 1) simplify the problem by ignoring the difficult components; 2) solve the problem for a "simple" system and extend the knowledge to a complex system; and 3) adopt a numerical approach. While these three approaches provide practical assessment for some cases, their applicability is limited due to their assumptions.

This paper is structured as follows: the first section lists three approaches, assumptions, and limitations; the second section formulates the problem in a different coordinate system than the polar coordinate system (PCS), and discusses the differences between the two; the third section lists the solution process to solve the reconstructed problem; the fourth section presents the analytical solution; the fifth section shows a set of examples; and the sixth section lists the conclusions and future studies.

# Approaches for solving the swing equation

## Coupled oscillator model

If the complexity associated with power systems is ignored, there might be a problem such as the swing equation. The first approach is to ignore the complexity of power systems, and to apply the knowledge from a different domain of science. It was found that in an oscillatory motion, if the frequency is spread more than the coupling between the oscillators, each oscillator runs at its own frequency. Otherwise, the system spontaneously maintains synchronization [6]. The field of synchronization in networks is reviewed in [7]. The Kuramoto model [8] provides an analytic function as follows:

$$\frac{d\delta_i^K}{dt} = \omega_i^K + \frac{\varsigma^K}{N}\sum_{j=1}^{N}\sin\left(\delta_i^K - \delta_j^K\right).$$

For the coupled oscillator model (COM), a system of particles aims to minimize the energy function

$E\left(\delta^K\right) = \sum_{\{i,j\}} a_{ij}\left[1-\cos\left(\delta_i^K - \delta_j^K\right)\right] - \sum_k \omega_k^K\delta_k^K \cong \frac{1}{2}\sum_{\{i,j\}} a_{ij}\left(\delta_i^K - \delta_j^K\right)^2 - \sum_k \omega_k^K\delta_k^K$. Term $E(\delta^K)$ features a phase-cohesive minimum with interacting particles no further than a certain angle $\gamma^K$. A solution has cohesive phases if there is an angle $\gamma^K$ between 0 and $\pi/2$ that is the maximum phase distance among all the pairs of connected oscillators, i.e., $|\delta_i - \delta_j| \le \gamma^K$ for a line connecting two nodes $i$ and $j$. In [9], it is demonstrated that the mechanical analogy applies to the electric power grid to yield the phase cohesiveness. This approach involves a very light computation cost, but the applicability may be limited due to the difference between a generic network and the electric power grids.

## Lyapunov stability

The second approach is to solve the equation for a simple problem without ignoring the nonlinear feature of the equation. To construct a simple problem, we consider a system with a single machine and infinite bus where the loss is ignored ($\gamma_{ik}$ = 0). In physics, we often define zero potential at an infinitely remote location, so that there is no influence from the existing fields. Similarly, we define an infinite bus so that the terminal voltages do not vary with the internal voltage angle (and therefore with time). However, the terminal voltages may change with the network condition. Then, the nonlinear component $\left|\tilde{y}_{ik}\right| E_i E_k \sin\left(-\theta_k + \delta_i - \gamma_{ik}\right)$ is rewritten as $p_{\max}\sin\delta_i$, where $p_{\max}\left(=\left|\tilde{y}_{ik}\right| E_i E_k\right)$ is a constant dependent on the network condition. There are three different conditions regarding the condition for a disturbance: pre-fault, post-fault, and on-fault (See Fig 1).

Fig 1 shows that the corresponding $p_{max}$ decreases in the order of pre-fault, post-fault, and on-fault conditions ($p_{\max}^{pre-fault} > p_{\max}^{post-fault} > p_{\max}^{on-fault}$). In the pre-fault condition, the operation point is determined where the sine curve intersects with a horizontal line $p_i^{mech} - g_{ii}E_i^2$ (Point $a$ in Fig 1 (B), note that losses are ignored, i.e., $g_{ii}$ = 0). At $a$, the generator is synchronized with the system frequency. During a fault, the sine function drops down the on-fault sine curve, but the internal voltage angle remains unchanged ($\delta_{pre}$), because the angle must be continuous over time ($a \rightarrow b$). Therefore, the electric power output is less than the mechanical power input, and the difference must be stored in the rotor in terms of mechanical energy to meet the energy balance. The internal voltage angle increases accordingly ($b \rightarrow c$), which makes the rate of the internal voltage angle deviate from the system frequency. The fault will shortly be monitored and cleared, and the sine curve follows the post-fault curve that the internal voltage angle follows ($c \rightarrow d$). Similarly, the internal voltage moves to a new curve at $\delta_{clear}$ in which the mechanical power input is less than the electric power output. The angle will increase further until the stored energy is exhausted at $\delta_p$ ($d \rightarrow e$). The integration of the left-hand side between two synchronized points of the swing equation (after ignoring the damping term) yields zero (Area 1 = Area 2). This is called the equal-area criterion. At $\delta_p$, the generator recovers the synchronization with the system frequency, but it cannot maintain the operation point because the electric output is greater than the mechanical power input and the stored energy is already exhausted. The rotor angle decreases until $\delta_n$, which is set by the equal-area curve (Area 3 = Area 4) and the rotor, swings between the two angles [ $\delta_n$, $\delta_p$] if no damping exists. With the damping, the swing range decreases and settles down to $\delta_{post}$, where the mechanical power input and electric output are balanced.

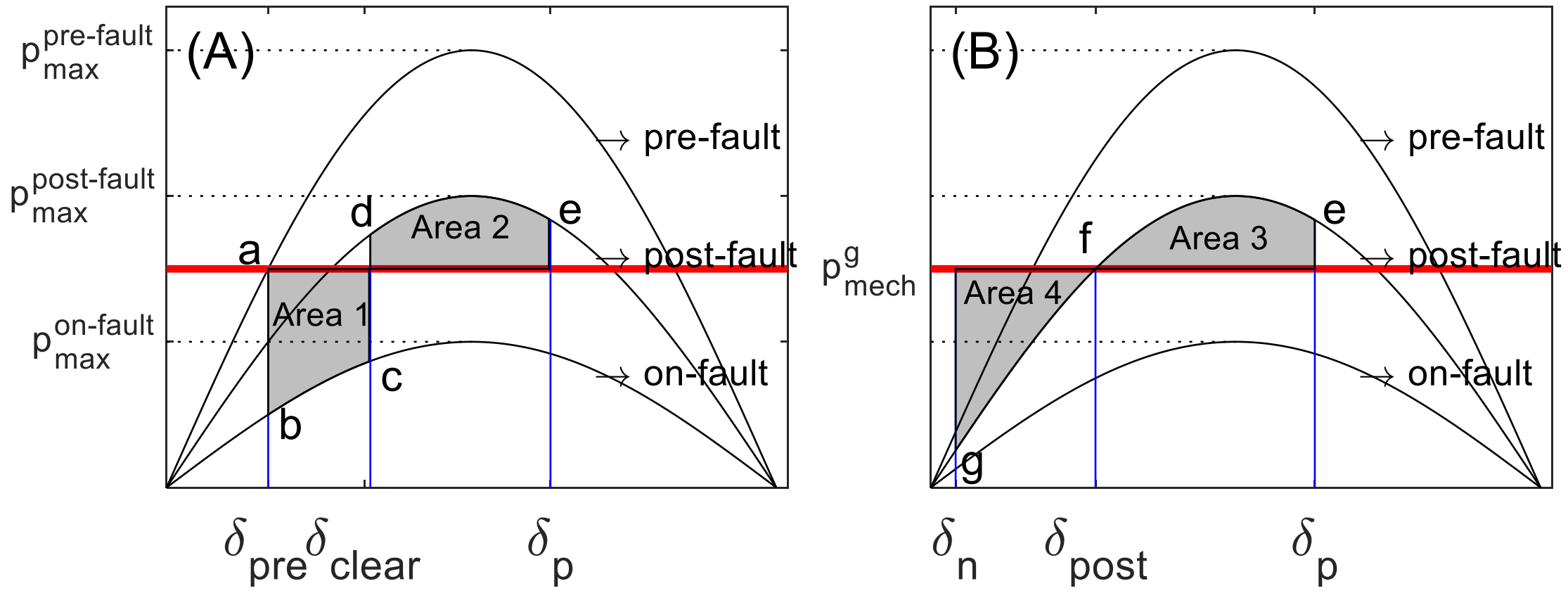

**Fig 1.** *p-δ* curve to illustrate the equal-area criterion.

Unfortunately, this equal-area criterion cannot generally be extended to a large-scale network with multiple machines. According to Lyapunov stability in nonlinear dynamic theory, one can tell if a system will be stable in the future if a Lyapunov function is identified to meet some conditions [10]. While there is no rule to construct a Lyapunov function, DM is developed under various assumptions [10 – 17] that often includes a lossless network (zero damping), no consideration of reactive power, and constant voltage magnitudes that are difficult to justify physically. Under these physically unjustifiable assumptions, the system is stable if one can find a Lyapunov function. On the other hand, the failure to find such a function (or the nonexistence of the function) does not necessarily mean the system will be unstable. Rather, the stability would be undetermined. Even though the advantage of this approach involves relatively inexpensive computation costs, it yields a stability region subset in addition to the physically infeasible assumptions.

Both the COM and the DM ignore the losses and variations of the voltage magnitudes. Further, because both assumptions are not physically justifiable, the application of the approaches is either highly limited or a conservative application is performed.

## Numerical approach

The third approach is to accommodate all the details of the swing equation and Kirchhoff's laws, and conduct a numerical computation. Often, a numerical integration or differentiation approach is applied to solve nonlinear differential equations and recent advancements in fast computation makes the numerical approach an attractive option [18], [19]. For the swing equation, a numerical method (such as the Euler or Runge–Kutta method) is used to find the temporal changes of internal and terminal voltages [20], [21]. The advantage of this approach is the capability to solve any complex function. However, in addition to the high computation costs, the numerical method involves a truncation error. For example, the errors associated with the Euler and modified Euler methods are in the order of $\vartheta\left(\|\Delta x\|^2\right)$ [20], and that of the Runge–Kutta method is $\vartheta\left(\|\Delta x\|^4\right)$ [22]. If there is a subset of modes that diverge exponentially, but in which the magnitudes are initially very small,

the modes may be underestimated. The numerical approach also requires precise estimation of the model parameters. It should be noted that numerical simulation did not capture the impact of the 1996 WSCC (Western System Coordinating Council) system outage [23].

# Write down the problem

## Feynman's algorithm

According to Dr. Robert J. Thomas (an Emeritus Professor at Cornell University), Feynman's algorithm for solving hard problems is a process of thinking about problems in a different way. The swing equation, $M_i \frac{d^2\delta_i}{dt^2} + D_i \frac{d\delta_i}{dt} = p_i^{mech} - g_{ii}E_i^2 - |\tilde{y}_{ik}| E_i E_k \sin(-\theta_k + \delta_i - \gamma_{ik})$, is formulated in the PCS because the equation describes the angular motion of a rotor inside a generator. The PCS provides a concise form of the equation of the motion; the differential terms are multiplied with scalar constants that result in a linear form for the left-hand side; and on the right-hand side, the magnitudes of internal voltage and terminal voltage are linear. However, the nonlinearity involved in voltage angles makes the swing equation difficult to solve. In this paper, instead of the conventional approach using the PCS, we construct the problem in the Cartesian coordinate system (CCS) and aim to solve it without physically unjustifiable assumptions.

## Observations in transient studies

We begin with a list of facts that are observed in transient studies.

O1. *E* (internal voltage of a generator) at *Ibus* (see Section "Network flows" for its definition) is constant [24], which is a general assumption applied to the transient studies in power systems,

$$O1 = \left\| \begin{bmatrix} 1 - \sqrt{x_i^2 + y_i^2} \Big/ E_i \cdots & \frac{1}{E_i}\sqrt{2x_i\left(\frac{dx_i}{dt}\right) + 2y_i\left(\frac{dy_i}{dt}\right)} \cdots \end{bmatrix}^T \right\|_F .$$

O2. O2 = $|O_i(t) - O_i^{0+}|$ where $O_i(t) = \left(-q_i + b_{ii}E_i^2\right)\big/ M_i + \left(d\delta_i/dt\right)^2$; $b_{ii}$ is the imaginary part of admittance; the superscript *0+* represents "immediately after contingency"; and $t > t_0$, i.e., the sum of the reactive power injection and the rotor speed changes slowly over time strictly after a disturbance occurs before losing synchronization. The square of the rotor speed $\left(d\delta_i/dt\right)^2$, the square of the rate to deviate from the synchronization, is in general smaller than $\left(-q_i + b_{ii}E_i^2\right)\big/ M_i$ while maintaining synchronization. According to O2, while maintaining synchronization, $|O_i(t)|$ is a slowly varying function that follows the Karamata representation theorem [25]: $|O_i(t)| = \left| \frac{-q_i + b_{ii}E_i^2}{M_i} + \left(\frac{d\delta_i}{dt}\right)^2 \right| = \exp\left[\eta(t) + \int_{t_1}^{t} \frac{\mu(y)}{y} dy\right]$ where

$$\begin{cases} \lim_{t\to\infty} \eta(t) = \eta_\infty \\ \lim_{t\to\infty} \mu(t) = 0 \end{cases}$$

for $t \geq t_1$ ($> t_0$). The variation of $O_i(t)$ over a short time period is negligible so that $O_i(t)$ = $\left(-q_i + b_{ii}E_i^2\right)\big/ M_i + \left(d\delta_i/dt\right)^2 \approx \left(-q_i^0 + b_{ii}E_i^2\right)\big/ M_i + \left(d\delta_i/dt\right)^2\Big|_{t=0}$ = $O_i(t_0)$. O2 is defined as $|O_i(t) - O_i(t_0)|$.

When a disturbance occurs it is possible for some generators to stray slightly off perfect synchronization. However, the rate of deviation from synchronization is always kept small before losing synchronization. The impacts of these approximations are discussed in a later section.

## Variables

A natural choice of variables in CCS would be the real and the imaginary components of voltages. The choice leads to $E_iE_k\sin(-\theta_k+\delta_i-\gamma_{ik})=-\cos\gamma_{ik}(E_i\cos\delta_i)(E_k\sin\theta_k)+\cos\gamma_{ik}(E_i\sin\delta_i)(E_k\cos\theta_k)-\sin\gamma_{ik}(E_i\cos\delta_i)(E_k\cos\theta_k)$ $+\sin\gamma_{ik}(E_i\sin\delta_i)(E_k\sin\theta_k)$. A better choice is to rotate the voltage angles by the phase angle of the line between the internal and terminal buses ( $\delta_i-\gamma_{ik}$), and the choice yields only two terms by $E_iE_k\sin(-\theta_k+\delta_i-\gamma_{ik})=-x_iv_y^k+y_iv_x^k$ where $x_i=E_i\cos(\delta_i-\gamma_{ik})$, $y_i=E_i\sin(\delta_i-\gamma_{ik})$, $v_x^k=E_k\cos\theta_k$, and $v_y^k=E_k\sin\theta_k$. For CCS, two variables are necessary ($x$ and $y$) instead of one ($\delta$) to describe the angular motion of internal voltages, and an additional equation is imposed to preserve the constant internal voltage magnitudes $x_i^2+y_i^2=E_i^2$.

## Load modeling

The development of a new load model is beyond the scope of this study. Instead, we attempt to integrate the existing load models for our proposed formulation. A widely used load model is outlined in [26]. Loads located at a same bus are integrated into a single load, and this load is separated and modeled into four subgroups: (Category I) an induction load, (Category II) a frequency-dependent or a time-dependent load, (Category III) a load with constant impedance or a load with constant current, and (Category IV) the remaining loads. While Categories I and II are integrated in the swing equation, the admittance matrix $Y$ is modified to incorporate Categories III and IV in terms of the voltage–current relationship.

**Category I: an induction load.** This has nonzero inertia, meaning it appears in the swing equation as a synchronous machine that remains in the swing equation. The load is modeled similarly to that of a generator, or more specifically, to that of a negative generator.

**Category II: a frequency dependent or a time dependent load.** A frequency dependent load is $d_j=d_j^0+m_j(d\delta_j/dt)$ [8], where the first component $d_j^0$ represents a load with a fixed impedance, and the second term $m_j(d\delta_j/dt)$is the frequency-dependent load. If a synchronous machine is located at the same bus as the load, the coefficient of the first time-derivative term is the sum of both the damping and frequency-dependent terms. Similarly, a time dependent load is $d_j=m_j(dd_j/dt)+d_j^0$ [27]. For this type of load, the constant term is integrated into Category IV.

**Category III: a load with constant impedance or with constant current.** These are integrated in $i_l=i_l^{cc}+y_l^{ci}v_l$, where $i_l^{cc}$ and $y_l^{ci}$ model the constant current and the constant impedance at Bus $l$, respectively. This expression makes it possible to convert these loads into the diagonal shunt element in the admittance matrix $Y_{bus}$ ($y_l$) or a constant in $i_L=diag(y_L^{ci})v_L+i_L^{cc}$.

**Category IV: a remaining load.** The remaining load introduces nonlinear characteristics, and it cannot characterize load characteristics on a constant voltage node in the system due to non-dependency on the voltage angle at the node. The BIG model in [28] and [29] is an attempt to integrate the load as a linear model (similar to loads with constant impedance or current), and show a good match for static loads. Interested readers may wish to read a literature survey on modeling loads in [30]. The type of load may be interpreted with a Taylor series expansion in terms of voltages near the operation point:

$I_l = \frac{s_l^*}{v_l^*} = \frac{\left(i_l^0 v_l + \bar{y}_i |v_l|^2\right)^*}{v_l^*} = i_l^0 + \bar{y}_i^* v_l$. Combined with the load modeling constant current and constant impendence, the current is expressed in terms of voltage $i_L = i_L^0 + diag\left(y_L^0\right) v_L$.

## Network flows

To establish a set of equations, an electric power grid is redefined, and a set of buses is introduced to model the internal voltage of a synchronous machine, *Ibus*; in other words, a *Kbus* is directly connected to an *Ibus*. Note that synchronous machines include induction motors, frequency- and time-dependent loads, and synchronous generators (but not asynchronous renewable generators). No *Ibus* is connected to multiple *Kbuses,* while a *Kbus* can be connected to multiple *Ibuses*. If two *Kbuses* are directly connected, an *Mbus* is inserted between them. Fig 2 illustrates the definitions of the *Ibus*, *Kbus*, and *Mbus*. Fig 2 (A) shows the one-line diagram of a modified IEEE 3-bus system with three generators, and Fig 2 (B) is the one-line diagram of the system modeled in this study. The generators are modeled as *Ibuses* (in red); the top two buses where the generators are directly connected are modeled as *Kbuses* (in green); and the bottom bus that is not connected to a generator is modeled as an *Mbus* (in blue), respectively. To prevent two *Kbuses* from being directly connected, another *Mbus* (vertical bus in blue) is inserted. Because all the buses in the original network remain in the model, and because *NI* of *Ibuses* and additional *Mbus* are introduced, the number of buses for this network model is always greater than that for the original network. It should be emphasized that unlike DM, in which the network reduction is applied to make the problem simple, this study preserves the topology of the network.

Suppose a disturbance occurs between $t = 0$ and $t = 0^+$ (immediately after the disturbance). During the disturbance, the angular motion of the physical rotor must be continuous; that is, instantaneous change in the angular motion is zero. Unlike rotor angles, voltages do change abruptly, while Kirchhoff's laws are obeyed both before and after the disturbance. Using Kirchhoff's laws, the voltages are computed to meet the real and the reactive power balances at *Kbus* and *Mbus* immediately after the disturbance. After the disturbance, the angular motion of the rotors and voltages at all the buses correspondingly adjusts to meet both the swing equations and Kirchhoff's laws. The admittance matrix of the redefined grid that accommodates Categories III and IV loads is in a block structure associated with both *Ibuses* and *Mbuses* because there is no direct connection between them. For the *Kbuses* and the *Mbuses* in the network model shown in Fig 2 (B), Ohm's law finds $i = Yv$ where the currents $i$ and the voltages $v$ at the buses are also expressed in terms of loads, i.e., $i_{KM} = i_{KM}^0 + diag\left(y_{KM}\right) v_{KM}$.

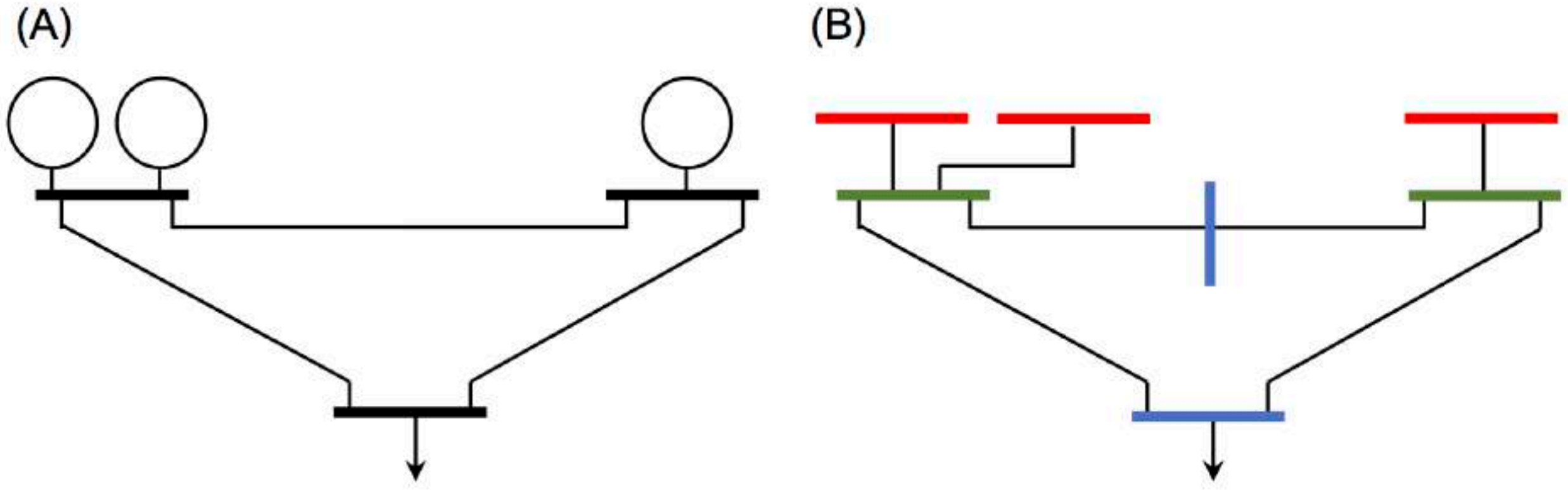

**Fig 2.** Schematic one-line diagrams of (A) the IEEE 3-bus system and of (B) the system with three *Ibuses* (in red), two *Kbuses* (in green), and two *Mbuses* (in blue).

https://doi.org/10.1371/journal.pone.0225097.g002

Note that induction, frequency-dependent, and time-dependent loads are not included in the current–voltage relationship. With no direct connection between *Ibus* and *Mbus*, Ohm's law over the network modeled in this work (Fig 2 (B)) is $\begin{pmatrix} i_{Ibus} \\ i_{Kbus} \\ i_{Mbus} \end{pmatrix} = \begin{bmatrix} Y_I^I & Y_I^K & 0 \\ Y_K^I & Y_K^K & Y_K^M \\ 0 & Y_M^K & Y_M^M \end{bmatrix} \begin{pmatrix} v_I \\ v_K \\ v_M \end{pmatrix}$. Because an *Ibus* is only connected to one *Kbus*, the inclusion of the buses is a radial extension of the original network. Therefore, the currents on the *Kbus* and the *Mbus* at a given voltage must be invariant to the choice of network model:

$$\begin{pmatrix} i_{Kbus} \\ i_{Mbus} \end{pmatrix} = \begin{bmatrix} Y_K^I & Y_K^K & Y_K^M \\ 0 & Y_M^K & Y_M^M \end{bmatrix} \begin{pmatrix} v_I \\ v_K \\ v_M \end{pmatrix} = Y_{KM} \begin{pmatrix} v_K \\ v_M \end{pmatrix} + \begin{pmatrix} i_{Kbus}^0 \\ i_{Mbus}^0 \end{pmatrix} \tag{1}$$

The parameters $i_{Kbus}^0$ and $i_{Mbus}^0$ represent the constant current and constant impedance components, respectively. Note that at *Kbuses* and *Mbuses*, only loads (including zero loads) exists because the loads in this study excludes the induction, frequency-dependent, and time-dependent loads. Eq (1) yields the following relationship in terms of modified voltages at *Ibus* ($w_I$):

$$\begin{pmatrix} v_K \\ v_M \end{pmatrix} = \left\{ Y_{KM} - \begin{bmatrix} Y_K^K & Y_K^M \\ Y_M^K & Y_M^M \end{bmatrix} \right\}^{-1} \left[ \begin{pmatrix} Y_K^I \\ 0 \end{pmatrix} R w_I - \begin{pmatrix} i_{Kbus}^0 \\ i_{Mbus}^0 \end{pmatrix} \right] = \begin{bmatrix} H_{KI} \\ H_{MI} \end{bmatrix} w_I + \begin{pmatrix} v_K^I \\ v_M^I \end{pmatrix} \tag{2}$$

where $R = \begin{bmatrix} \cos\gamma_{ik} & \sin\gamma_{ik} \\ -\sin\gamma_{ik} & \cos\gamma_{ik} \end{bmatrix}; v_I = Rw_I; w_I = \begin{pmatrix} x_I \\ y_I \end{pmatrix}$. Using (2), the real and the imaginary components of the bus voltages are

$$v_x^k = e_k^T v_K = e_k^T H_{KI} w_I + e_k^T v_K^I,\; v_y^k = e_{k+NK}^T v_K = e_{k+NK}^T H_{KI} w_I + e_{k+NK}^T v_K^I \tag{3}$$

## Revisiting swing equation

Typically, swing equations are written in the PCS, as this is convenient for describing angular motion. The PCS inevitably introduces the exponential function (i.e., sinusoidal function) to express the real power injection, which makes it difficult to solve analytically. In a DC power flow model [26], we often linearize the sinusoidal function by taking $\sin\theta \approx \theta$ and $\cos\theta \approx 1$ as $\theta \approx 0$. However, in the swing equation, δ to describe the angular motion is not small enough to apply the approximation. The power injection from *Ibus i* to *Kbus k* is as follows:

$$\begin{aligned} p_{i\to k}^{elec} &= g_{ii}E_i^2 + |\tilde{y}_{ik}| E_i E_k \sin(-\theta_k + \delta_i - \gamma_{ik}) = g_{ii}E_i^2 - |\tilde{y}_{ik}| x_i v_y^k + |\tilde{y}_{ik}| y_i v_x^k \\ q_{i\to k}^{elec} &= b_{ii}E_i^2 - |\tilde{y}_{ik}| (x_i v_x^k + y_i v_y^k) \to |\tilde{y}_{ik}| (x_i v_x^k + y_i v_y^k) = -q_{i\to k}^{elec} + b_{ii}E_i^2 \end{aligned} \tag{4}$$

The definitions of $x_i$ and $y_i$ lead to

$$\frac{d\delta_i}{dt} = -\frac{1}{y_i}\frac{dx_i}{dt} = \frac{1}{x_i}\frac{dy_i}{dt} = \frac{1}{E_i^2}\left( x_i \frac{dy_i}{dt} - y_i \frac{dx_i}{dt} \right) \text{ and } \frac{d^2\delta_i}{dt^2} = \frac{1}{E_i^2}\left( x_i \frac{d^2 y_i}{dt^2} - y_i \frac{d^2 x_i}{dt^2} \right) \tag{5}$$

With (4) and (5), the swing equation in CCS becomes

$$y_i\left( M_i \frac{d^2 x_i}{dt^2} + D_i \frac{dx_i}{dt} \right) - x_i \left( M_i \frac{d^2 y_i}{dt^2} + D_i \frac{dy_i}{dt} \right) + E_i^2 \left( p_{i\to k}^{mech} - g_{ii}E_i^2 - |\tilde{y}_{ik}| v_x^k y_i + |\tilde{y}_{ik}| v_y^k x_i \right) = 0 \tag{6}$$

In comparison to the swing equation in the PCS, (6) is a heterogeneous nonlinear second-order differential equation with multi-variables; the time derivatives are multiplied with the variables; and the last term is nonlinear. Additionally, the conditions related to the constant internal voltage magnitude and to its derivatives are also nonlinear. If one combines the conditions with (6), the resulting equation becomes highly complex. Instead, we derive

$$x_i\left(M_i\frac{d^2x_i}{dt^2}+D_i\frac{dx_i}{dt}\right)+y_i\left(M_i\frac{d^2y_i}{dt^2}+D_i\frac{dy_i}{dt}\right)+E_i^2\left[M_i\left(\frac{d\delta_i}{dt}\right)^2\right]=0 \quad (7)$$

Equations (6) and (7) lead to

$$\begin{aligned}
&M_i\frac{d^2x_i}{dt^2}+D_i\frac{dx_i}{dt}+\left[-q_i+b_{ii}E_i^2+M_i\left(\frac{d\delta_i}{dt}\right)^2\right]x_i+\left(p_i^{mech}-g_{ii}E_i^2\right)y_i-|\tilde{y}_{ik}|E_i^2v_x^k=0\\
&M_i\frac{d^2y_i}{dt^2}+D_i\frac{dy_i}{dt}-\left(p_i^{mech}-g_{ii}E_i^2\right)x_i+\left[-q_i+b_{ii}E_i^2+M_i\left(\frac{d\delta_i}{dt}\right)^2\right]y_i-|\tilde{y}_{ik}|E_i^2v_y^k=0\\
&s.t.\, x_i^2+y_i^2=E_i^2;\, x_i\frac{dx_i}{dt}+y_i\frac{dy_i}{dt}=0
\end{aligned} \quad (8)$$

Note that (8) involves no approximation. It is interesting that the reactive power injection appears in (8), while the swing equation considers only the real power balance. The constraints are conditions that the differential equations hold under.

From O2, $\left[\left(-q_i+b_{ii}E_i^2\right)/M_i+\left(d\delta_i/dt\right)^2\right]-\left[\left(-q_i^{0+}+b_{ii}E_i^2\right)/M_i+\left(d\delta_i/dt\right)^2\Big|_{t=0^+}\right]$ is approximately zero, or simply $O_i(t)$ = $O_i(0+) = O_i^0$:

$$\begin{aligned}
&M_i\frac{d^2x_i}{dt^2}+D_i\frac{dx_i}{dt}+M_iO_i^{0+}x_i+\left(p_i^{mech}-g_{ii}E_i^2\right)y_i-|\tilde{y}_{ik}|E_i^2v_x^k=0\\
&M_i\frac{d^2y_i}{dt^2}+D_i\frac{dy_i}{dt}-\left(p_i^{mech}-g_{ii}E_i^2\right)x_i+M_iO_i^{0+}y_i-|\tilde{y}_{ik}|E_i^2v_y^k=0
\end{aligned} \quad (9)$$

Furthermore, the voltages at *Kbus* are expressed in terms of $w_I$ as shown in (2)

$$\begin{aligned}
&e_i^T\frac{d^2w_I}{dt^2}+\frac{D_i}{M_i}e_i^T\frac{dw_I}{dt}+\left(O_i^{0+}e_i^T+\frac{p_i^{mech}-g_{ii}E_i^2}{M_i}e_{NI+i}^T-\frac{|\tilde{y}_{ik}|E_i^2}{M_i}e_k^TH_{KI}\right)w_I-\frac{|\tilde{y}_{ik}|E_i^2}{M_i}\left(e_k^Tv_K^I\right)=0\\
&e_{NI+i}^T\frac{d^2w_I}{dt^2}+\frac{D_i}{M_i}e_{NI+i}^T\frac{dw_I}{dt}+\left(-\frac{p_i^{mech}-g_{ii}E_i^2}{M_i}e_i^T+O_i^{0+}e_{NI+i}^T-\frac{|\tilde{y}_{ik}|E_i^2}{M_i}e_{NK+k}^TH_{KI}\right)w_I-\frac{|\tilde{y}_{ik}|E_i^2}{M_i}\left(e_{NK+k}^Tv_K^I\right)=0\\
&s.t.\, O1^2=\left[\frac{1}{E_i^2}vec\left(e_ie_i^T+e_{i+NI}e_{i+NI}^T\right)\right]^T\left[\begin{matrix}w_I\otimes w_I & 2w_I\otimes\frac{dw_I}{dt}\end{matrix}\right]-\left(\begin{matrix}1 & 0\end{matrix}\right)=0;\, O2=O_I(t)-O_i^{0+}=0
\end{aligned} \quad (10)$$

$O1^2$ represents the element-wise square. If Bus *i* is directly connected to a frequency- and/or a time-dependent load, the damping coefficient is modified to accommodate the frequency- and/or time-dependent constant $m_i$; that is, $D_i^{new} = D_i + m_i$. The nodal swing equation with only frequency- and/or time-dependent load without synchronized machine with O2 becomes [8]

$$\begin{aligned}
&e_i^T\frac{d\tilde{w}_I}{dt}+\left(\frac{M_{ref}}{D_i}O_i^{0+}e_i^T+\frac{p_i^{mech}-g_{ii}E_i^2}{D_i}e_{NI+i}^T-\frac{|\tilde{y}_{ik}|E_i^2}{D_i}e_k^TH_{KI}\right)\tilde{w}_I-\frac{|\tilde{y}_{ik}|E_i^2}{D_i}\left(e_k^Tv_K^I\right)=0\\
&e_{NI+i}^T\frac{d\tilde{w}_I}{dt}+\left(-\frac{p_i^{mech}-g_{ii}E_i^2}{D_i}e_i^T+\frac{M_{ref}}{D_i}O_i^{0+}e_{NI+i}^T-\frac{|\tilde{y}_{ik}|E_i^2}{D_i}e_{NK+k}^TH_{KI}\right)\tilde{w}_I-\frac{|\tilde{y}_{ik}|E_i^2}{D_i}\left(e_{NK+k}^Tv_K^I\right)=0
\end{aligned} \quad (11)$$

Note that $R_I$ associated with $w_I$ is an identity matrix. To distinguish the modified voltages derived by the load from a physical internal voltage, $w_I$ is introduced in (11). The buses where the frequency- or time-dependent loads are located belong to the *Ibus*. For a system with *NI* of *Ibuses* and *NK* of *Kbuses*, (10) is further simplified with the constraints (O1 and O2 are small) as follows:

$$\frac{d^2 w_I}{dt^2} + diag\left(\frac{D_I}{M_I}\right)\frac{dw_I}{dt} + Lw_I + l = 0 \tag{12}$$

where $e_i^T L = O_i^{0+} I + \frac{p_i^{mech} - g_{ii}E_i^2}{M_i} e_i e_{i+NI}^T - \frac{|\tilde{y}_{ik}|E_i^2}{M_i} e_i e_k^T H_{KI}$, $e_{i+NI}^T L = -\frac{p_i^{mech} - g_{ii}E_i^2}{M_i} e_{i+NI} e_i^T + O_i^{0+} I - \frac{|\tilde{y}_{ik}|E_i^2}{M_i} e_{i+NI} e_{k+NK}^T H_{KI}$,

$e_i^T l = -\frac{|\tilde{y}_{ik}|E_i^2}{M_i}\left(e_k^T v_K^I\right)$, $e_{i+NI}^T l = -\frac{|\tilde{y}_{ik}|E_i^2}{M_i}\left(e_{k+NK}^T v_K^I\right)$, $e_i^T w_I = x_i$, $e_{i+NI}^T w_I = y_i$, and the superscript *T* refers to transpose. Similarly, (11) becomes

$$\frac{d\tilde{w}_I}{dt} + \tilde{L}\tilde{w}_I + \tilde{l} = 0 \tag{13}$$

where $e_i^T \tilde{L} = \frac{M_{ref}}{D_i} O_i^{0+} I + \frac{p_i^{mech} - g_{ii}E_i^2}{D_i} e_i e_{i+NI}^T - \frac{|\tilde{y}_{ik}|E_i^2}{D_i} e_i e_k^T H_{KI}$, $e_{i+NI}^T L = -\frac{p_i^{mech} - g_{ii}E_i^2}{D_i} e_{i+NI} e_i^T + \frac{M_{ref}}{D_i} O_i^{0+} I$ $-\frac{|\tilde{y}_{ik}|E_i^2}{D_i} e_{i+NI} e_{k+NK}^T H_{KI}$, $e_i^T \tilde{l} = -\frac{|\tilde{y}_{ik}|E_i^2}{D_i}\left(e_k^T v_K^I\right)$, $e_{i+NI}^T \tilde{l} = -\frac{|\tilde{y}_{ik}|E_i^2}{D_i}\left(e_{k+NK}^T v_K^I\right)$, $e_i^T \tilde{w}_I = \tilde{x}_i$, $e_{i+NI}^T \tilde{w}_I = \tilde{y}_i$, and $x_i$, $y_i$ are the modified voltages derived from the frequency- or time-dependent load at Bus *i*. Let *z* be [$w_I$; $dw_I/dt$; $\tilde{w}_I$], and then with the constraints (O1 and O2 are small) (12) and (13) becomes

$$\frac{dz}{dt} = Tz + b \quad \text{where } T = \begin{bmatrix} 0 & I & 0 \\ -L & -diag\left(D_I/M_I\right) & 0 \\ 0 & 0 & -\tilde{L} \end{bmatrix} \text{ and } b = \begin{pmatrix} 0 \\ -l \\ -l \end{pmatrix} \text{ s.t. } \begin{cases} x_i^2 + y_i^2 = E_i^2 \\ x_i \frac{dx_i}{dt} + y_i \frac{dy_i}{dt} = 0 \end{cases} \tag{14}$$

(14) is a constrained homogeneous linear first-order differential equation with multi-variables. There is no known solution to the constrained differential equation. Fig 3 illustrates our approach to solving the swing equation. The arrows with solid lines refer equivalent, and those with dashed lines represent similar in the proximity proportional to the relaxation applied. S1 is the solution to the conventional swing equation formulated in PCS, which is identical to S2 in CCS and S3 because (8) is equivalent to (10) with the constraint of O2 = 0. Equality constraints are identical to the inequality constraints if the upper bounds are zeros, i.e., S3 = S4. S5 is the solution to the problem relaxed by $\vartheta(\Delta_E)$. Note that $t_E + \delta t$ is the first time when any of the relaxed constraints is binding, and that $\delta t$ is strictly positive. In the range of [0, $t_E$], no constraints are binding, which makes S6 involve nonbinding constraints. In the optimization theory, the solution to a constrained problem equals that of the same problem without the constraints that are not binding. Therefore, S7 is the same as that to S6 even though no constraints are taken into consideration in solving S7. The process finds S7 is in the proximity of S1 by $\vartheta(\Delta_E)$ if $\delta t \ll t_E$ and $\Delta_E$ is kept small enough to ensure the validity of the solution. The justification is outlined in the section describing the validity region with projections so that the error remains small. This concludes the first step of Feynman's algorithm, *write down the problem.*

## Think hard: solve the swing equation

### Initial Condition

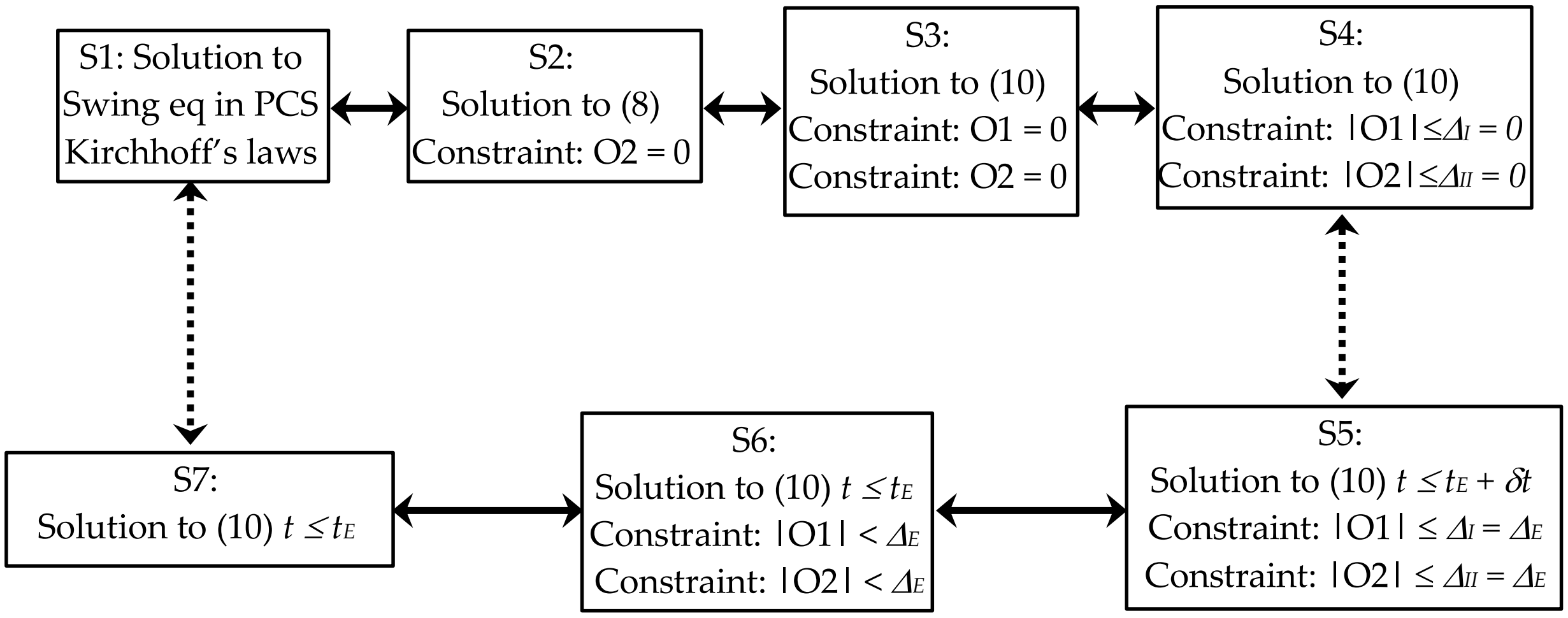

**Fig 3. Solutions to various forms of swing equation and its relaxations.**
https://doi.org/10.1371/journal.pone.0225097.g003

To solve a first-order differential equation, a set of initial conditions are required. $\delta$ represents a physical angular motion that is continuous in time; i.e., $\delta_i^{0+} = \delta_i^0$. When a disturbance occurs, the mechanical power input does not change instantaneously, but the electric power output $p_{i\to k}^{elec}$ exhibits a sudden but finite change. The change is represented by a step function $f\left(x \in R^1\right) = \sum_i m_i \chi_{A_i}(x)$ where $m_i$ is a finite real number and $\chi_A$ is an indicator function over an interval $A$, $\chi_A(x) = \begin{cases} 1 & \text{if } x \in A \\ 0 & \text{otherwise} \end{cases}$. Because the time interval between 0 and $0^+$ has a finite length, the Lebesgue integral of $p_{i\to k}^{elec}$ is $\int p_{i\to k}^{elec}\, dt = m\ell(A)$, where $\ell(A)$ is the length of the time interval between 0 and $0^+$, $\int_{t=0}^{t=0^+} p_{i\to k}^{elec}\, dt \simeq 0$. The value of $\omega_i$ (= $d\delta_i/dt$) immediately after the disturbance occurs is computed from the integration of the swing equation over the time interval $\omega_i^{0+} = \omega_i^0$. Similarly, for the frequency- or time-dependent loads, $\tilde{\omega}_i^{0+} = \tilde{\omega}_i^0$. The rotor angle δ (in terms of time) is differentiable, and its first derivative ω is bounded; hence, the rotor angle $\delta$ is Lipschitz continuous over time.

## General solution to the swing equation

The general solution to (14) that satisfies $\frac{dz_G}{dt} = Tz_G$ is as follows:

$$z_G = \Phi u \text{ where } u = \begin{pmatrix} e^{\tilde{\lambda}_1 t} \\ \vdots \\ e^{\tilde{\lambda}_{2N} t} \end{pmatrix} \text{ and } \begin{pmatrix} \tilde{\lambda}_1 \\ \vdots \\ \tilde{\lambda}_{2N} \end{pmatrix} = eig(T) \tag{15}$$

where $\Phi$ is the coefficient matrix to be determined. The base solution $u$ may contain terms associated with complex eigenvalues, because $T$ is generally an asymmetric real-valued matrix. For any complex eigenvalue, its complex conjugate is also a complex eigenvalue. Let such a pair be $\lambda_l + \lambda_{l+1}\sqrt{-1}$ and $\lambda_l - \lambda_{l+1}\sqrt{-1}$. Then, the $l^{th}$ corresponding pair in the base solution is

$$\begin{pmatrix} u_l \\ u_{l+1} \end{pmatrix} = \begin{bmatrix} e^{\lambda_l t}\cos(\lambda_{l+1} t) \\ e^{\lambda_l t}\sin(\lambda_{l+1} t) \end{bmatrix} \tag{16}$$

It is possible to represent $u$ in terms of a real-valued function as follows:

$$u = \begin{pmatrix} u_{\mathrm{Re}} \\ u_{Co} \end{pmatrix} \text{ where } u_{\mathrm{Re}} = \begin{pmatrix} u_{\mathrm{Re}}^{1} \\ \vdots \\ u_{\mathrm{Re}}^{N_{\mathrm{Re}}} \end{pmatrix},\ u_{Co} = \begin{bmatrix} u_{Co}^{[1,2]} \\ \vdots \\ u_{Co}^{\{N_{Co}-1,N_{Co}\}} \end{bmatrix},\ u_{\mathrm{Re}}^{n} = e^{\lambda_{\mathrm{Re}}^{n} t}, \text{ and } u_{Co}^{\{l,l+1\}} = \begin{pmatrix} e^{\lambda_{Co}^{\{l\}} t}\cos\left(\lambda_{Co}^{\{l+1\}} t\right) \\ e^{\lambda_{Co}^{\{l\}} t}\sin\left(\lambda_{Co}^{\{l+1\}} t\right) \end{pmatrix} \tag{17}$$

where $n$ and $l$ are the indices of real and complex eigenvalues, respectively. Differentiating $u$ with respect to time introduces a block-diagonal matrix $D_l$ such that $du/dt = D_l u$, where $D_i = \begin{bmatrix} diag(\lambda_{\mathrm{Re}}) & 0 \\ 0 & bldiag\left(\begin{bmatrix} \lambda_{Co}^{\mathrm{Re}} & -\lambda_{Co}^{\mathrm{Im}} \\ \lambda_{Co}^{\mathrm{Im}} & \lambda_{Co}^{\mathrm{Re}} \end{bmatrix}\right) \end{bmatrix}$, $\lambda_{Re}$ is the real eigenvalues, $\lambda_{Co}^{\mathrm{Re}}$ and $\lambda_{Co}^{\mathrm{Im}}$ are real and imaginary components of the complex eigenvalues of $T$, respectively. It is noteworthy that $D_l$ is obtained from the block Schur decomposition [31] $T = U_T^{-1} D_l U_T$, so that $D_l$ has a block-diagonal structure. Therefore, $T$ and $D_l$ are related through the similarity transformation. In (15), $\Phi$ is the time-invariant coefficient matrix, and the differential equation yields

$$\frac{dz_G}{dt} = \frac{d}{dt}(\Phi u) = \Phi\frac{du}{dt} = \Phi D_l u = T\Phi u \tag{18}$$

Because $u$ is not a null vector, (18) introduces a Sylvester equation; that is, $\Phi D_l - T\Phi = 0$. The Bartels–Stewart algorithm is an efficient way to solve a Sylvester equation [31], and its computational cost is $\mathcal{O}(N^3)$. It involves the orthogonal reduction of $D_l$ and $T$ matrices into triangular form using the $QR$ factorization, and then solving the resulting triangular system via back-substitution. Because $D_l$ is a block-diagonal matrix comprised of the eigenvalues of $T$, their eigen-spaces overlap. Therefore, the Bartels–Stewart algorithm is not applicable to solve (18). Consideration of the null space yields $vec(\Phi) \in null\left[\left(I \otimes T - D_\lambda^T \otimes I\right)^T\right]$, where $\otimes$ represents the Kronecker product. Therefore, the QR-factorization of $\left(I \otimes T - D_\lambda^T \otimes I\right)$ reveals the space of $\Phi$. Let $\psi_l$ and $\Psi_l$ be the $l^{\text{th}}$ vector spanning the null space of $\left(I \otimes T - D_\lambda^T \otimes I\right)$ and the matrix form of the vector, respectively. Then, $\Phi$ is a linear combination of all $\Psi$s.

$$z_G = \left(\sum_l \beta_l \Psi_l\right) u, \text{ where } \Psi_l D_l - T\Psi_l = 0 \tag{19}$$

## Solution to meet initial conditions

We claim that the following function is the solution to (14):

$$\hat{z}(t) = z_G + \begin{pmatrix} -L^{-1} l \\ 0 \\ -\tilde{L}^{-1}\tilde{l} \end{pmatrix} = \sum_{l=1}^{4NI+2ND} \beta_l \Psi_l u + \begin{pmatrix} -L^{-1} l \\ 0 \\ -\tilde{L}^{-1}\tilde{l} \end{pmatrix} \tag{20}$$

It is straightforward to prove this claim as follows:

$$\frac{d\hat{z}(t)}{dt} = \sum_{l=1}^{4NI+2ND} \beta_l \Psi_l \frac{du}{dt} = \sum_{l=1}^{4NI+2ND} \beta_l \Psi_l D_\lambda u = T \sum_{j=1}^{4NI+2ND} \beta_l \Psi_l u = T\left[\hat{z}(t) - \begin{pmatrix} -L^{-1}l \\ 0 \\ -\tilde{L}^{-1}\tilde{l} \end{pmatrix}\right] = T\hat{z}(t) + \begin{pmatrix} 0 \\ -l \\ -\tilde{l} \end{pmatrix} = T\hat{z}(t) + b \tag{21}$$

Using $\omega_i^{0+} = \omega_i^0$ and $\tilde{\omega}_i^{0+} = \tilde{\omega}_i^0$, the initial condition of $\hat{z}(t=0^+)$ is

$$w_I\Big|_{t=0^+} = w_I^0, \left.\frac{dw_I}{dt}\right|_{t=0} = -Jw_I^0, \text{ and } \left.\frac{d\tilde{w}_I}{dt}\right|_{t=0} = -\tilde{L}\tilde{w}_I^0 - \tilde{l} \quad \text{where } J = \begin{bmatrix} 0 & diag(\omega_i^{0+}) \\ -diag(\omega_i^{0+}) & 0 \end{bmatrix} \tag{22}$$

With a set of vectors $\xi$, $\xi_l = \Psi_l u(t=0)$, $\beta$ is determined from the following least square process:

$$\beta^* = \arg\min_\beta \left\| \left[\xi_1 \cdots \xi_{4NI} \cdots \xi_{4NI+2ND}\right]\beta - \begin{pmatrix} w_I^0 + L^{-1}l \\ -Jw_I^0 \\ -\tilde{L}\tilde{w}_I^0 - \tilde{l} \end{pmatrix} \right\|_2 \tag{23}$$

The analytical solution to the swing equation that satisfies Kirchhoff's laws is

$$w_I = \Theta u - L^{-1}l;\ v_K = H_{KI}\Theta u - H_{KI}L^{-1}l + v_K^I;\ v_M = H_{MI}\Theta u - H_{MI}L^{-1}l + v_M^I \text{ where } \Theta = \sum_{r=1}^{4NI} \beta_r^* \left(\left[I\ 0\ 0\right]\Psi_r\right) \tag{24}$$

If all the real parts of the eigenvalues are negative, $u$ converges to zero as time increases. Therefore, under the circumstance and when O1 and O2 remain sufficiently small,

$w_I \to -L^{-1}l,\ \tilde{w}_I \to -\tilde{L}^{-1}\tilde{l},\ v_K \to -H_{KI}L^{-1}l + v_K^I, v_M \to -H_{MI}L^{-1}l + v_M^I$.

## Write down the solution

### Computational complexity

There are several steps to determining the complexity: finding voltages immediately after the disturbance, decomposing eigenvalues of $T$, and computing $\beta$. The first process has the same complexity as solving power flow equations $\vartheta(Nb^{1.5})$ [32]; the second and last processes involve $\vartheta[(NI+ND)^3]$ [31], because the dimension of $T$ is $(4NI+2ND) \times (4NI+2ND)$. Therefore, the first process is predominant if $Nb > (NI+ND)^2$. Even though additional computations are necessary at $\tau$, the first process does not need to be solved because the voltages are readily available from (2). Therefore, the additional computation may not necessarily increase the computational cost significantly.

### Stability assessment

The conventional stability assessment based on COM or DM certifies when a state is eventually stable. We propose another stability assessment—that the eigenvalues of $T$ play a key role.

***Type I***: stable if the real parts of all the eigenvalues are non-positive or $\Theta = 0$ (no disturbance).

***Type II***: operationally stable if the largest positive real eigenvalue $\lambda_m$ is small enough such that $1/\lambda_m \le T_{op}$, where $T_{op}$ is a time scale where transient stability is concerned. Typical time scales of transient stability studies are between sub-seconds to tens of seconds [33].

***Type III***: operationally stable if the coefficients corresponding to the terms in the base solution with positive real eigenvalues are small enough for the system to remain stable within $T_{op}$.

***Type IV***: unstable if the system divulges rapidly.

In this assessment scheme, the positive real parts of eigenvalues and the corresponding coefficients play a key role. Apart from *Type III* (in the presence of a disturbance), assessment is possible with the eigenvalues of the $T$ matrix. If no positive eigenvalues exist, the state is eventually stable.

## Validity region

The general solution of (20) to (10) is obtained without considering the constraints. It would be ideal to solve the ordinary differential equation (10) with the constraints–DAE. DAE integrates the differential variables and the algebraic variables. While DAE is complete, to the best of our knowledge, no analytical solution has currently been identified. Therefore, the approach in this study is to identify the validity region where the constraints hold. To discuss the change of the constraints (drift-off phenomena), if we do not explicitly consider them while solving the differential equation (such as P7 in Fig 3), suppose we want to solve the following pendulum problem, $dx/dt = u; du/dt = -\lambda x; dy/dt = v; dv/dt = -\lambda y - 1;$ $x^2 + y^2 - 1 = 0$. One may eliminate λ, and find a differential equation as follows: $\frac{u'}{x} + \frac{u^2+v^2-y}{x^2+y^2} = 0; \frac{v'+1}{y} + \frac{u^2+v^2-y}{x^2+y^2} = 0$ with the following constraints: $x^2 + y^2 = 1$ and $xu + yv = 0$. The numerical solution to the differential equation is evaluated without considering the constraints explicitly. Fig 4 (A) shows the drift-off phenomena of the constraints, and the errors associated with $x^2 + y^2 = 1$ and $xu + yv = 0$ grow quadratically and linearly, respectively [34].

Eq (10) is analogous to constrained mechanical systems that can incorporate the repeated projection of a numerical solution onto the solution manifold; projections on position constraints and on velocity constraints for improving the stability (see Fig 4 (B)). The error remains negligible with well-developed projections; hence, the solution (20) to (10) is valid without consideration of the constraints. However, the projections are not a linear process, which makes it difficult to integrate them into (10). Instead, we define a validity region where solution (20) is valid, and the projections are made at the boundary of the validity region to compensate for the drift-off. The induced errors and the approach to compensate the errors are discussed in the following section.

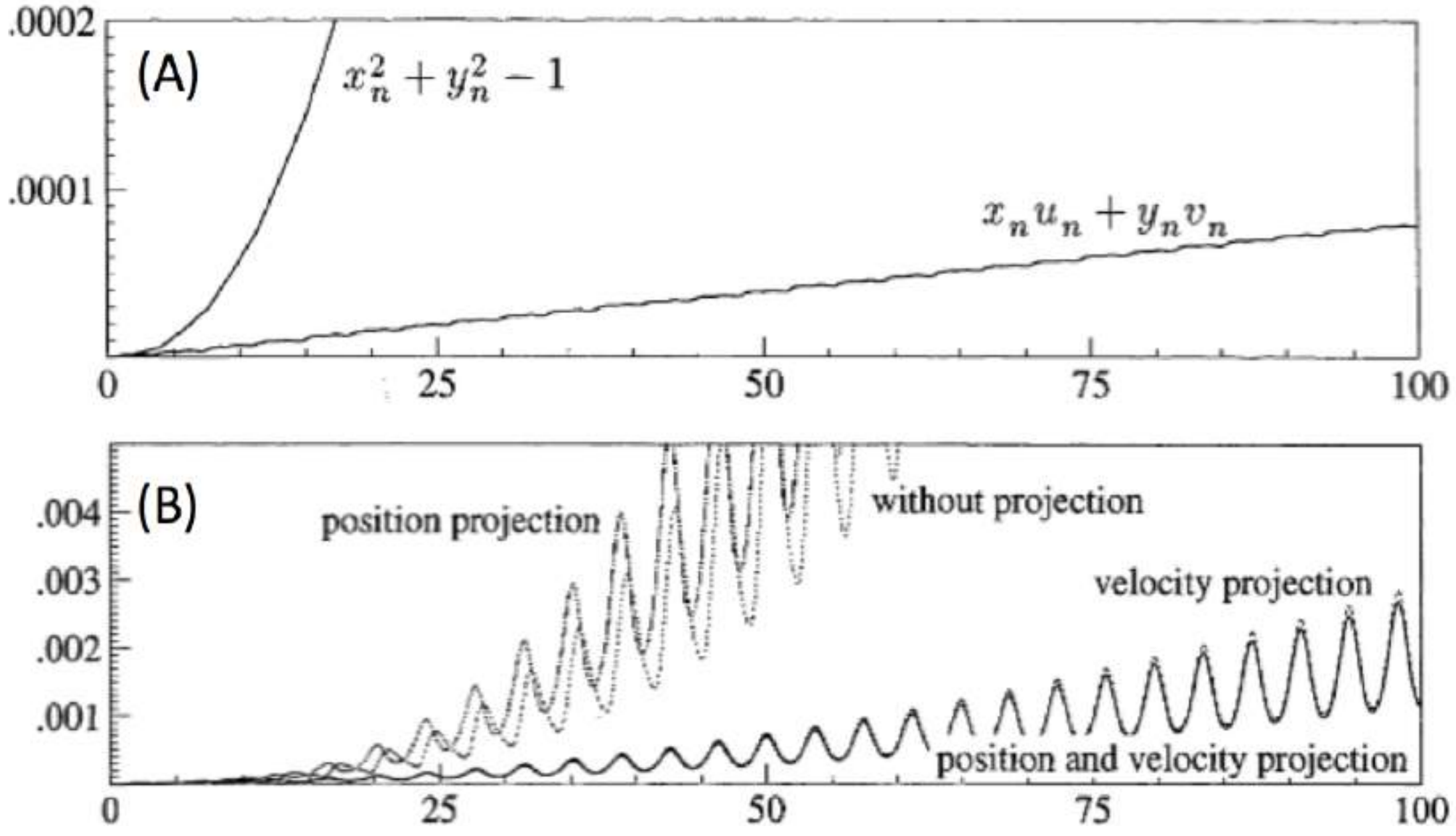

**Fig 4.** (A) Errors in the constraints and (B) global error with various projections. Both figures are from Ref. [34].

## Internal voltages, $E_I$

Suppose, in a time range of [0, $\Delta t$], the internal voltages deviate from the nominal values significantly, which affects $z$ in (20), meaning the constraints do not hold. Because (20) yields the analytical expression of $x_I$ and $y_I$ in terms of $t$, one can evaluate a projection vector $\Delta_{con}$ that includes the error in $E_I$ (position projection, $\breve{p}_E^i = 1 - \sqrt{x_i^2 + y_i^2} / E_i$) and the error in the first derivative (velocity projection, $\breve{v}_E^i = \sqrt{2x_i(dx_i/dt) + 2y_i(dy_i/dt)} / E_i$), i.e., $\Delta_{con} = \left\| \left( \breve{p}_E^T \;\; \breve{v}_E^T \right)^T \right\|_2$. The constraints hold if the 2-norm of the projection vector does not exceed a threshold $\Delta_E$.

## Error in $T$

The matrix $T$ in (10) is exact except for one element each row by an approximation that $O_i(t) = \left(-q_i + b_{ii}E_i^2\right)/M_i + \left(d\delta_i/dt\right)^2 = \left(-q_i^0 + b_{ii}E_i^2\right)/M_i + \left(d\delta_i/dt\right)^2\Big|_{t=0} = O_i^{0+}$. There might be a discrepancy between the true $z$ and the estimated $z$ due to the nonnegligible O2. The impact will affect two components: (1) the inaccuracy in estimating $T$ and the eigenvalues of $T$, and (2) the error in the coefficients according to $\Phi$, $\delta\Phi$.

The impact of O2 in the inaccuracy is as follows: When the change in $O_i(t)$ is in the order of $\varepsilon_T$, the impact propagates to $T$, $D_l$, and correspondingly $\Phi$ and $\Psi$. The change in $T$ due to $O_i(t)$ affects the eigenvalues of $T$ that are in the block-diagonal in $D_l$. Suppose $\lambda$ is a simple eigenvalue of $T$ and $u_L$ and $u_R$ are the corresponding left- and right- eigenvectors. $\vartheta(\varepsilon_T)$ changes in $T$ can induce $\varepsilon_T/s(\lambda)$ in the eigenvalues [31],

$$\left|\lambda - \hat{\lambda}\right| \approx \frac{\varepsilon_T}{s(\lambda)} \text{ where } U_L^H T U_R = diag(\lambda_1, \cdots, \lambda_N),\; s(\lambda) = \left|u_L^H u_R\right|, \text{ and } \varepsilon_T = \|\delta T\|_F / \|T\|_F \tag{25}$$

Eq (25) indicates that the impact of the change in $O_i(t)$ is a linear change in the eigenvalues of $T$, the block-diagonal elements in $D_l$. Let $\hat{T}$ and $\hat{D}_l$ be the true matrices; $\delta T$ and $\delta D_l$ be the difference matrices between the true and the approximated matrices (i.e., $\hat{T} = T + \delta T$ and $\hat{D}_l = D_l + \delta D_l$); and $\hat{\Phi}$ and $\delta\Phi$ be the true solution matrix corresponding to $\hat{T}$ and $\hat{D}_l$ and the difference matrix, respectively. Further, define $\rho$ and $\sigma$: $\rho = 2\left[\max\left(\|T\|_F, \|D_l\|_F\right)\right] \left\| \left(T \otimes I - I \otimes D_l^T\right)^{-1} \right\|_2$ and $\|\delta T\|_F \le \sigma \|T\|_F$, $\|\delta D_l\|_F \le \sigma \|D_l\|_F$. $D_l$ is a similarity transformed matrix of $T$ (i.e., $D_l = P^{-1}TP$); hence, $\|D_l\|_F = \|P^{-1}TP\|_F \le \kappa_F(P)\|T\|_F$. From $\Phi D_l - \Phi T = 0$ and $\hat{\Phi}\hat{D}_l - \hat{T}\hat{\Phi} = 0$, one finds

$$\hat{\Phi}\hat{D}_l - \hat{T}\hat{\Phi} \approx \Phi\delta D_l + \delta\Phi D_l - T\delta\Phi - \delta T\Phi = 0 \rightarrow \delta T\Phi - \Phi\delta D_l = T\delta\Phi - \delta\Phi D_l \tag{26}$$

Eq (26) can be rewritten in terms of the Kronecker product,

$$\delta T\Phi - \Phi\delta D_l = \left(\delta T \otimes I - I \otimes \delta D_l^T\right) vec(\Phi) \text{ and } T\delta\Phi - \delta\Phi D_l = \left(T \otimes I - I \otimes D_l^T\right) vec(\delta\Phi) \tag{27}$$

Using the Kronecker product, one finds the upper/lower bounds of the Frobenius norm of a product between two matrices $A$ and $B$ as follows:

$$\|B\|_F = \|A^{-1}AB\|_F = \|vec(A^{-1}AB)\|_2 \le \|I \otimes A^{-1}\|_2 \|vec(AB)\|_2 = \|A^{-1}\|_2 \|AB\|_F \rightarrow \|AB\|_F \ge \|A^{-1}\|_2^{-1} \|B\|_F \tag{28}$$

Eqs (27) and (28), and the triangle inequality theorem yield

$$\begin{aligned} \|\delta T\Phi - \Phi\delta D_l\|_F &= \left\| \left(\delta T \otimes I - I \otimes \delta D_l^T\right) vec(\Phi) \right\|_F \le \left\|\delta T \otimes I - I \otimes \delta D_l^T\right\|_2 \|\Phi\|_F \\ \|T\delta\Phi - \delta\Phi D_l\|_F &= \left\| \left(T \otimes I - I \otimes D_l^T\right) vec(\delta\Phi) \right\|_F \ge \left\| \left(T \otimes I - I \otimes D_l^T\right)^{-1} \right\|_2^{-1} \|\delta\Phi\|_F \end{aligned} \tag{29}$$

and using the triangular inequality and the fact that the Frobenius norm is no less than 2-norm [31]:

$$\frac{\|\delta\Phi\|_F}{\|\Phi\|_F} \le \left\|\delta T \otimes I - I \otimes \delta D_l^T\right\|_2 \left\|\left(T \otimes I - I \otimes D_l^T\right)^{-1}\right\|_2 \le \left(\|\delta T\|_F + \|\delta D_l^T\|_F\right)\left\|\left(T \otimes I - I \otimes D_l^T\right)^{-1}\right\|_2 \tag{30}$$

With the definition of $\rho$ and $\sigma$, (30) becomes:

$$\frac{\|\delta\Phi\|_F}{\|\Phi\|_F} \le \rho\sigma \sim \vartheta(\varepsilon) \tag{31}$$

Because a time-varying $O_i(t)$ determines $\varepsilon$, one can compute the $\tau_\varepsilon$ that the relative error in $T$ remain a threshold $\Delta_T$, i.e., $\varepsilon = \|\delta T\|_F / \|T\|_F \le \Delta_T$ for all $t \le \tau_\varepsilon$. Let $y_z$ be the first derivative of $z$ with respect to time, then (14) yields: $y_z = Tz_0 + b$ where $y_z = dz/dt|_{t=0}$ and $z = z_0 + ydt$. In the time range, there might be an error in $T$ involving the error that propagates in $y_z$ and $z$ as follows:

$$\delta z \cong \delta z_0 + \left(T\delta z_0 + \delta T z_0\right)\Delta t = \left(I + T\Delta t\right)\delta z_0 + \left(\Delta t \delta T\right) z_0 \tag{32}$$

## Beyond the validity region

The boundary of the validity region is defined either when the constraints do not hold, or when the error in $T$ exceeds the threshold. At the boundary of the validity region, the values for $z$ and $T$ are updated as described. With the updated values, (14) still holds because the errors in (14) and the constraints remain less than the threshold. Therefore, the problem in (14) will be solved with the initial condition that is the updated values for $z$. It is necessary to ensure consistency among the initial values; hence, a consistent initialization problem. This problem has been studied widely; (1) a small artificial step with the backward Euler method [35], [36], (2) Taylor series expansion [37], [38], and (3) graph theoretic algorithm to obtain the minimal set of equations for differentiation to solve for consistent initial values [39], [40].

Eqs (12) and (13) and the constraints are rewritten as follows:

$$\begin{gathered} f\left(t,u_I,w_I,du_I/dt\right) = \begin{cases} \dfrac{du_1}{dt} + diag\left(\dfrac{D_I}{M_I}\right)u_1 + Lw_I + l = 0 \\ \dfrac{du_2}{dt} + \tilde{L}u_2 + \tilde{l} = 0 \end{cases} \rightarrow \frac{du_I}{dt} + \hat{D}_I u_I + \hat{L}_I w_I + \hat{l}_L = 0 \\ g\left(t,u_I,w_I\right) = \begin{cases} \left[vec\left(e_i e_i^T + e_{i+NI} e_{i+NI}^T\right)\right]^T \left(w_I \otimes w_I\right) - E_i^2 = 0 \\ \left[vec\left(e_i e_i^T + e_{i+NI} e_{i+NI}^T\right)\right]^T \left(w_I \otimes u_I\right) = 0 \end{cases} \\ \rightarrow F\left(t,u_I,w_I,du_I/dt\right) = \begin{bmatrix} f\left(t,u_I,w_I,du_I/dt\right) \\ g\left(t,u_I,w_I\right) \end{bmatrix} = 0 \\ \text{where } u_I = \frac{d\omega_I}{dt}; u_2 = \tilde{w}_I; u = \begin{pmatrix} u_1^{2NI} \\ u_2^{2ND} \end{pmatrix}; \hat{D}_I = \begin{bmatrix} diag\left(\dfrac{D_I}{M_I}\right) & 0 \\ 0 & \tilde{L} \end{bmatrix}; \hat{L}_I = \begin{bmatrix} L \\ 0 \end{bmatrix}; \hat{l}_I = \begin{pmatrix} l \\ \tilde{l} \end{pmatrix} \end{gathered} \tag{33}$$

The cardinalities of $f$ and $g$ are $2NI+2ND$ and $2NI$, respectively. By the definition of the variables in (33), $u$ and $s$ are classified as differential variables and algebraic variables. Accordingly (33) is called a DAE. At a given $w_I^0$ and $u_I^0$, the Newton-Raphson method finds:

$$\begin{bmatrix} \Delta f\left(t,u_I,w_I,du_I/dt\right) \\ \Delta g\left(t,u_I,w_I\right) \end{bmatrix} = \begin{bmatrix} hI & \hat{L}_I & \hat{D}_I \\ 0 & \hat{M}_w & \hat{M}_u \end{bmatrix} \begin{pmatrix} \Delta u_I'/h \\ \Delta w_I \\ \Delta u_I \end{pmatrix} \rightarrow \begin{pmatrix} \Delta u_I'/h \\ \Delta w_I \\ \Delta u_I \end{pmatrix} = \begin{bmatrix} hI & \hat{L}_I & \hat{D}_I \\ 0 & \hat{M}_w & \hat{M}_u \end{bmatrix}^{-1} \begin{pmatrix} \Delta f \\ \Delta g \end{pmatrix} \tag{34}$$

where $\hat{M}_w = \begin{bmatrix} \begin{pmatrix} 2w_I^{0T} M_1 \\ \vdots \\ 2w_I^{0T} M_N \end{pmatrix}^T & \begin{pmatrix} 2u_I^{0T} M_1 \\ \vdots \\ 2u_I^{0T} M_N \end{pmatrix}^T \end{bmatrix}^T$, $\hat{M}_u = \begin{bmatrix} 0 & \begin{pmatrix} w_I^{0T} M_1 \\ \vdots \\ w_I^{0T} M_N \end{pmatrix}^T \end{bmatrix}^T$.

The term in (34) is compensated to yield a consistent initial point immediately beyond the validity region. With the consistent initial point and updated $T$, the analytical solution in (20) is identified. Most components inside the curly bracket in (34) are constant. Therefore, only a few visits to the boundaries of the validity limits does not increase the computation time significantly if efficient rank-update techniques are employed.

## Insight on the dynamics

As in (12), $T$ has a block structure, and it can be broken into two submatrices as follows:

$$T = \underbrace{\begin{bmatrix} 0 & I & 0 \\ -L_{sys} & -diag(D_I/M_I) & 0 \\ 0 & 0 & -\tilde{L}_{sys} \end{bmatrix}}_{T_{sys}} - \underbrace{\begin{bmatrix} 0 & 0 & 0 \\ L_{op} & 0 & 0 \\ 0 & 0 & \tilde{L}_{op} \end{bmatrix}}_{T_{op}=\sum(+)-\sum(-)} \quad (35)$$

The first term is invariant with the operation point of specific contingency, but the second term varies. We consider the eigenvalue decomposition of $T$ as consecutive applications of the eigenvalue update and down-date of $T_{op}$ from the eigenvalues of $T_{sys}$; that is, the sum of multiple rank-1 [41] or rank-2 [42] updates (+) and multiple down-dates (-). The impacts of $T_{op}$ are to shift the eigenvalues of $T_{sys}$, which is bounded by the Bauer-Fike theorem [31]: $\min_{\lambda\in\lambda(T_{sys})} |\lambda - \hat{\lambda}| \le \left(\|U_\Phi\|_p \|U_\Phi^{-1}\|_p\right)\|T_{op}\|_p$ where $U_\Phi^{-1} T_{sys} U_\Phi = diag(\hat{\lambda}_1,\cdots,\hat{\lambda}_N)$. If the diagonal elements in $T_{op}$ are all zeros (or negligible), the Gershgorin circle theorem can be applied [31]. The Gershgorin disk is isolated from the other disks so that a disk contains precisely one eigenvalue of $T$. Motivated readers may find the process in [43]. Note that $T_{sys}$ only depends on the system, and $T_{op}$ varies with the dynamics. The following section outlines the terms affecting the stability of the system.

**Reactive power**, $q_i^c$ in $q_i^c + b_{ii}E_i^2$. This term negatively affects $L_{con}$, shifting the eigenvalues to the right. Reactive power supports voltages, which makes it difficult for a system to settle into new voltages. However, the reactive power injection is very small in comparison to the $b_{ii}E_i^2$ term; therefore, its impact on the transient stability is highly limited.

**Unsettled mechanical power**, $p_i^{mech}$ in $p_i^{mech} - g_{ii}E_i^2$. These terms are associated with the mechanical power in $L$, and have the same magnitudes with opposite signs; hence, their impacts on shifting the real eigenvalues of $T$ cancel out. However, their impacts are on shifting the imaginary components of the eigenvalues. It is intuitively correct that unsettled mechanical power enhances the oscillating motion. In many cases, the resistance of an internal generator model is zero, meaning $g_{ii}$ is zero. Unlike the reactive power injection, $p_i^{mech}$ plays a key role in stability assessment. A synchronization condition based purely upon the power injections is proposed in [9].

**Inter-voltage sensitivity matrix**, $H_{KI}$. As the sensitivity of $H_{KI}$ increases in a positive direction, the voltages at *Kbus* swing tightly together with the generators. The negative sensitivities in $H_{KI}$ imply that the voltages at *Kbus* swing against the angular motion of rotors. They act as a dragging force to stabilize the system. The impact of the inter-voltage sensitivities is to shift in the same direction as the sign of $H_{KI}$.

**Damping coefficient**. The damping term appears in the lower diagonal in $T$. Because the damping terms are non-negative, the impact is to shift the eigenvalues to the left. Therefore, the damping terms help to stabilize the system against disturbance.

**Inertia.** The terms reactive power, unsettled mechanical power, voltage sensitivity $H_{KI}$, and damping are all scaled by the inertia constant $M_i$. Regardless of their signs, the amplitudes are normalized in terms of inertia. The impact of inertia becomes clear in TDS in a way that agrees with this observation.

**Loads.** Loads are classified into three categories: synchronized induction load, frequency- or time-dependent load, and remaining loads. The remaining loads affect the sensitivity matrix $H_{KI}$, which is a part of the $L,\ L$, and $T_{op}$ matrices. The frequency- and time-dependent loads, and the synchronized induction loads, are taken into consideration in the swing equation (14). It is noteworthy that the $T$ matrix has a block diagonal structure between the two loads; therefore, the Eigen-space of each block is independent so that their subspaces are orthogonal.

## Illustrative examples and discussion

Simulations are performed for various IEEE model systems (IEEE 9, 14, 30, and 118-bus systems). To compare the results with both a coupled oscillator model [9] and DM, which assume a lossless system ($\gamma = 0$ for all the lines) and constant voltage magnitudes, the resistance components are all ignored. A phase cohesiveness for synchronization is also introduced: $\left|\theta_i - \theta_j\right| \le m \in \left[0, \pi/2\right]$. Electric power is generated at *Ibus* and injected into the grids; therefore, the angles at *Ibus* ($\delta_I$) are greater than those at *Kbus*. ($\delta_i \ge \theta_k$). The power flow from *Ibus i* to *Kbus k* (power injection at *PV buses* of the original network) is proportional to $\sin\left(\delta_i - \theta_k\right)$[26]. Therefore, the phase cohesiveness is equivalent to the constraints imposed on the injection between two directly connected oscillators. We found that as the system becomes unstable, the maximum angle differences after the disturbance are significantly higher than those in the stable case. Even though the values for the dynamics are not listed in [9], the phase cohesiveness for synchronization finds the trend correctly for the disturbances we tested. Similarly, the synchronization condition is also checked, and it was found that there is a $\gamma^K$ appearing in [9] to satisfy $\left\|L^{\dagger} p\right\|_{\varepsilon,\infty} \le \sin\gamma^K$ for the chosen equilibrium points. Because there can be many equilibrium points, it can be difficult to analyze the stability region of each point [13]. For the sake of visual presentation, the simulation results on the IEEE 9 bus system are discussed in this paper. Fig 5 shows the one-line diagram of the system, and Table 1 lists the data relevant to this study. In the proposed network modeling in this study, there are three *Ibuses*, three *Kbuses*, and six *Mbuses*. The pre-fault power flows are computed using the unified method based on the Kronecker product [44], and the threshold values for $\Delta_T$ and $\Delta_E$ are 1% and 10%, respectively. The scaling factor $h$ in this study is 0.1. All the numerical computations are performed using a Mac pro with two 2.93 GHz 6-core Intel Xeon processors.

The results are summarized in Table 2. In the table, 10% $d_8$ in the first column refers to 10% loss of loads at Bus 8; Fig in the second column is the figures associated with the event at the first column; $\Delta\delta_{max}$ under coupled oscillator column is the synchronization condition proposed in [9] where the threshold for the system is 0.129; $\Delta V_{st}$ under the DM column is the stability margin that is defined in the Appendix. For some cases, the stability assessments are undetermined and shown as "-", because the certificate is not issued. Time under the time domain simulation (TDS) and proposed model columns represent the computation time for numerical computations.

### Loss of loads

Three sets of cases are performed to simulate the loss of load at Bus 8; no loss, 10% loss, and entire loss. Prior to the loss of load, the system was in the steady state condition that was identified using the power flow study. When no disturbance occurs, the system should stay in the same steady state at $t = 0$. Immediately after the loss of load, a new operating point is found by solving (2) with the updated loads and $w_I$ at the steady state, because the rotor angles do not change instantaneously.

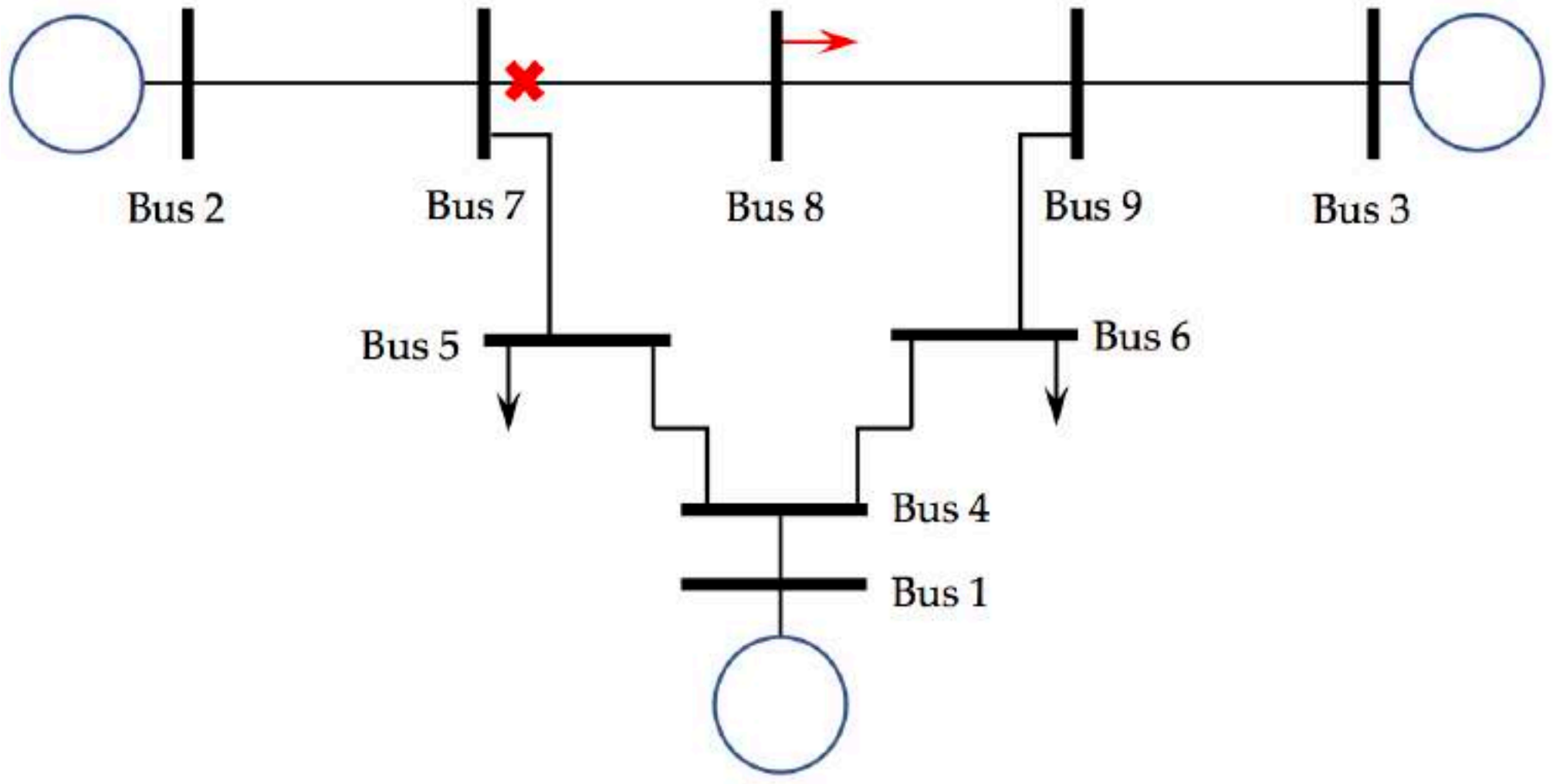

**Fig 5. IEEE 9-bus system.** The red arrow and red cross represent the loss of load and the line fault, respectively.

https://doi.org/10.1371/journal.pone.0225097.g005

**Table 1. Machine data for the IEEE 9-bus system illustrated in Fig 5.**

| Generator location | $M_i$ | $D_i$ | $g_{ii}$ | $b_{ii}$ | $\lvert E_i \rvert$ |
|---|---|---|---|---|---|
| Bus 1 | 2.364 | 0.0254 | 0.0 | -16.45 | 1.057 |
| Bus 2 | 0.640 | 0.0066 | 0.0 | -8.35 | 1.050 |
| Bus 3 | 0.301 | 0.0026 | 0.0 | -5.52 | 1.017 |

https://doi.org/10.1371/journal.pone.0225097.t001

**Table 2. Summary of the simulation results.**

| Event | Fig | Coupled oscillator | | DM | | TDS | | Proposed method | |
|---|---|---|---|---|---|---|---|---|---|
| | | $\Delta\delta_{max}$ | stability | $V_{margin}$ | stability | time | stability | time | stability |
| 10% $d_8$ | 6 | 0.116 | stable | 8.96 | stable | 3.80 | stable | 0.17 | stable |
| entire $d_8$ | 7 | 0.113 | stable | 9.21 | stable | 2.62 | stable | 0.55 | stable |
| on-fault | 8 | 0.137 | - | -34.87 | - | 3.45 | unstable | 0.11 | unstable |
| postfault | 10 | 0.155 | - | -1.98 | - | 3.37 | stable | 1.91 | stable |

https://doi.org/10.1371/journal.pone.0225097.t002

The stability of the post-fault operation point is estimated using DM and the synchronization condition proposed in [9]. The trajectory during the transient state is numerically calculated in terms of TDS. The details of TDS and of DM based on an energy function are outlined in the Appendix. For the case with no disturbance, even though the eigenvalues of $T$ are non-zeros, the coefficient $\Theta$ is zero (*Type I*) because the constant term in (20) is zero, meaning the system stays in the steady state at the pre-fault condition.

## Slight change in the load at Bus 8

Fig 6 illustrates the trajectories of (A) rotor angles $\delta$, of (B) voltage magnitudes $E$, of (C) $O_i(t)$, and of (D) computed magnitudes of the internal voltages when a 10% loss of the load at Bus 8 occurs at $t = 1$ s.

**Stability of the system.** In the post-fault state, the maximum angle differences in voltage angles immediately after the disturbance is 0.116 rad, while the synchronization condition reported from [9] is 0.129 rad. As shown Fig 6 (B), the variations of the voltage magnitudes are not significant, which indicates that a condition of both COM and DM holds. Because the maximum angle difference is less than the threshold, the stability assessment predicts the convergence to a stable state. The energy functions $V(\delta, \omega)$ for DM are evaluated at all equilibrium points to check the stability of the system. Based on to the stability margin of 8.96 ($= V_{cr} - V_{cl} > 0$), a stability certificate is issued by DM. TDS is also performed with a time step $\Delta t$ of 0.01 s. All three stability assessment approaches yield the same estimate—converging to a new equilibrium point.

The positive eigenvalues of $T$ are small, and the corresponding $\Theta$ is numerically negligible (*Type II*). As shown in Fig 6 (A), the rotor angles are all synchronous, and the voltage magnitudes quickly settle down to a new equilibrium point. The analytical solution and the numerical results from TDS are visually indistinguishable, as shown in Fig 6 (A). For clear presentation, the voltage magnitudes and the rotor speeds are generated by the proposed analytical approach in Fig 6 (B).

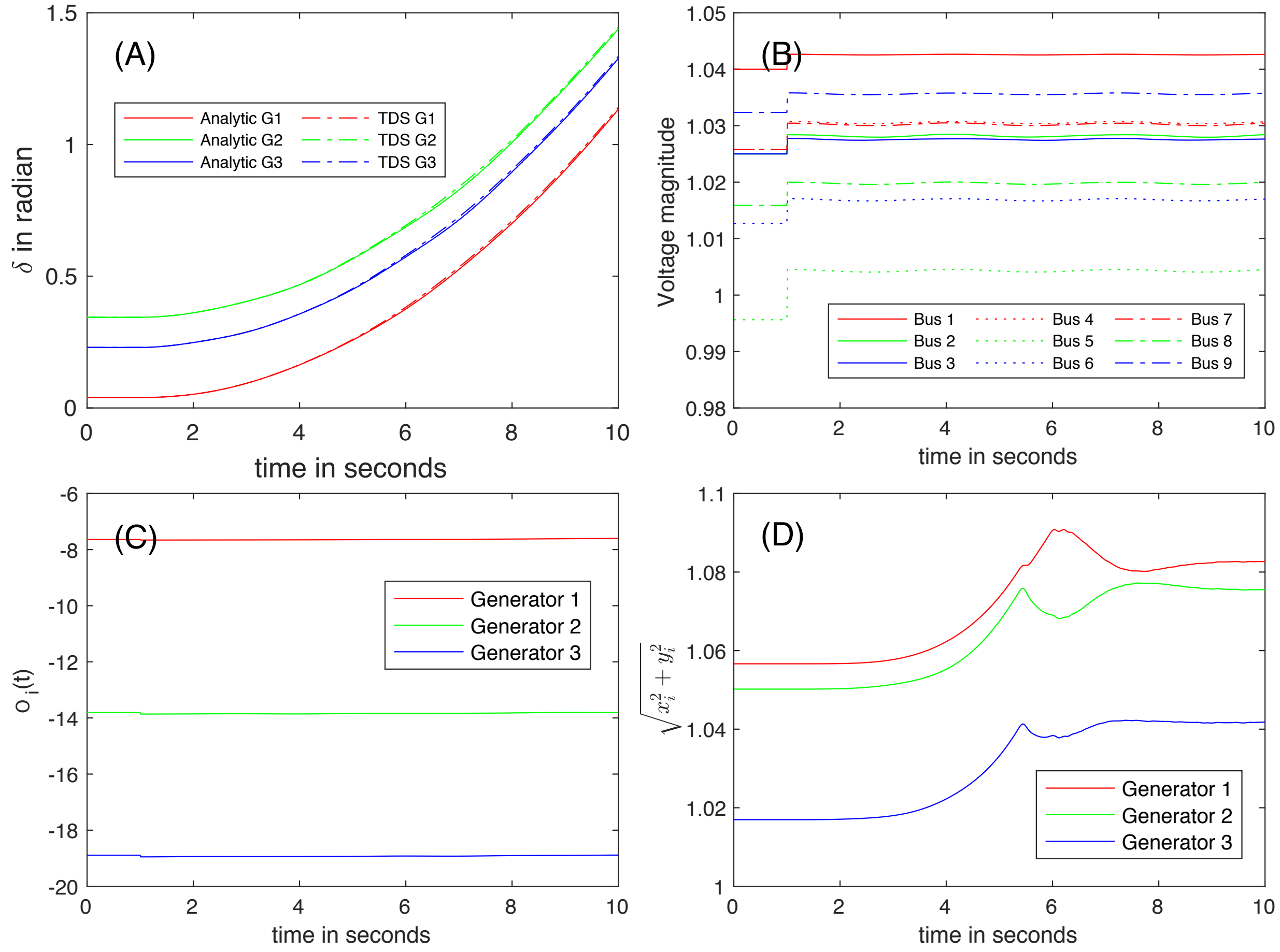

**Fig 6.** The trajectories of (A) $\delta$, of (B) $E$, of (C) $O_i(t)$, and of (D) the magnitudes of the internal voltages after 10% of loss of the load at Bus 8 (red arrow in Fig 5).

**Validity region.** In this study, the tolerance for ε $\left(=\|\delta T\|_F/\|T\|_F\right)$, $\Delta_T$, is set to 1% for all the numerical evaluations. The value of $\|T\|_F$ in the case of 10% loss of the load at Bus 8 is 16.71; hence, the tolerance of the impact of O2 toward $\|\delta T\|_F$ is $1.67 \times 10^{-2}$ (= 1% × 16.71). The variations of $O_i(t)$ are negligible, as shown in Fig 6 (C)—the value of $\|\delta T\|_F$ is $1.21 \times 10^{-4}$ ($\|\delta T\|_F/\|T\|_F = 7.24 \times 10^{-6}$) is $\vartheta(10^{-4})$. This means the impact of O2 is negligible, and the validity region extends to the entire time domain of the transient in this case. Fig 6 (C) indicates that $\eta(t)$ for the generators are constant after assuming $\mu(t) = 0$. Therefore, $|O_i(t)|$ follows the Karamata representation theorem (i.e., a slowly varying function), which also leads to the extended validity region. Fig 6 (D) shows that the estimated voltage magnitudes vary within the threshold ($\Delta_E$ = 10%) of the nominal voltages, and that O1 is less than the threshold most times. This implies only a few times of crossing the validity regions in 10 s. It is noteworthy to mention that in most simulations, O1 and O2 are small.

## Entire loss of the load at Bus 8

A significant change in the load at Bus 8 occurs at $t = 1$ s, and the generator dynamics are simulated. Fig 7 illustrates the trajectories of (A) $\delta$, (B) the voltage magnitudes at the terminal voltages, (C) $O_i(t)$, and (D) the estimated magnitudes of the internal voltages.

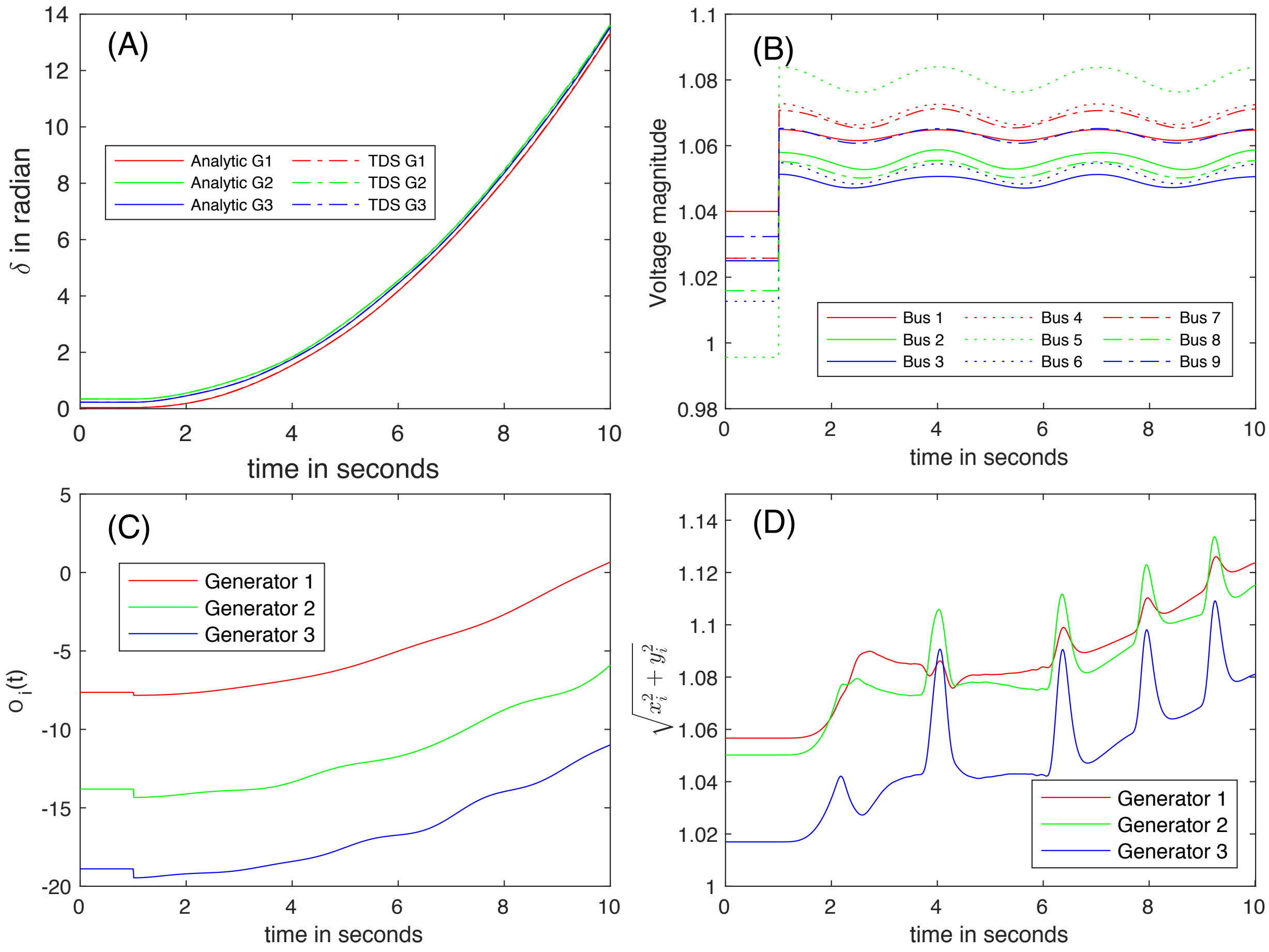

**Fig 7.** The trajectories of (A) $\delta$, of (B) $E$, of (C) $O_i(t)$, and of (D) the estimated magnitudes of internal voltages due to the entire loss of the load at Bus 8 at $t = 1$ sec (red arrow in Fig 5).

https://doi.org/10.1371/journal.pone.0225097.g007

**Stability of the system.** After the loss of load, the stability of the system is tested; (1) the maximum angle difference across the lines is 0.113 (<0.129), (2) the stability margin is 9.21 (>0), and (3) TDS with $\Delta t = 0.01$ s. All three methods find the stability of the system, and converge to a new equilibrium point after the disturbance. Due to the loss of loads at Bus 8, the reactive power injected to the grid exceeds the demands, and the uncompensated reactive power makes the voltages at the terminal buses increase (Fig 7 (B)). A condition of COM and DM that the terminal voltage magnitudes are constant holds marginally.

Fig 7 (A) shows the dynamics of rotor angle for the entire loss of the load at Bus 8, and exhibits no discrepancies between the analytic approach and TDS. The positive eigenvalues of $T$ are small, but the corresponding $\Theta$ is numerically negligible (*Type II*). Therefore, the proposed analytical solution also estimates the stability of the system correctly.

**Validity region.** Because the value of $\|T\|_F$ in the case of the entire loss of the load at Bus 8 is 17.61, the tolerance to the impact of O2 toward $\|\delta T\|_F$ is 0.18 (= 1% × 17.61). The variation in $O_i(t)$ of three generators over 10 s is approximately 8, and the corresponding $\|\delta T\|_F$ is $9.88 \times 10^{-2}$ ($\|\delta T\|_F / \|T\|_F = 5.61 \times 10^{-3}$). Therefore, the validity region extends to the entire time domain of the transient if O1 is small. As shown in Fig 7 (C), $\eta(t)$ (after assuming $\mu(t) = 0$) does not converge as $t$ goes to infinity; therefore, the Karamata representation theorem may not be applicable. Consistent with this observation, $O_i(t)$ is not slowly varying, but the impacts on $T$ are within the threshold. O1 often (5 peaks) reaches $\Delta_E$ (= 10%), as shown in Fig 7 (D). The projection method identifies the peaks (the boundaries of the validity regions), and the consistent initial points are evaluated using (34) for the application of (20), which increases the computation time.

## Line fault: on-fault trajectory

At $t = 1$ s, a three-phase fault occurs on the line connecting Buses 7 and 8 near Bus 7 (red cross mark shown in Fig 4). Using the pre-fault power flow solution by applying the unified power flow analysis approach in [44], the parameters to formulate the swing equations are identified. Due to the continuation of the rotational movement, the rotor angle $\delta$ remain continuous in time regardless of the change in the network. The element in $Y_{bus}$ corresponding to Bus 7 is increased to represent a high admittance to ground, and the voltage at Bus 7 collapses. With this modification, the on-fault voltages and the parameters for the swing equations in the on-fault trajectory are computed. Fig 8 illustrates the on-fault trajectories of (A) $\delta$, of (B) $E$, of (C) $O_i(t)$, and of (D) the estimated magnitudes of the internal voltages when the fault is not cleared.

**Stability of the system.** After the disturbance, Bus 2 is isolated from the system that is directly connected to Generator 2, and its voltage magnitude reduces to zero (Fig 8 (B)). Because no load is located at Bus 2, the electric power injection becomes zero, and the electrical power input from the generator is stored in the rotor in the form of mechanical power (increased rotor speed). For the rest of the system, the change in the power supply by the isolated Bus 2 (effectively the loss of Generator 2) is compensated by the other generators. This leads to a change in rotor angle, as shown in Fig 8 (A). Clearly, the reactive power injection changes abruptly (see Fig 8 (C)).

The maximum angle difference for COM is 0.137 rad, while the value for $\gamma$ in [9] is 0.129, which does not meet the synchronization condition. If the method fails to issue a certificate, the stability of the system is undetermined. However, the short-circuit makes Bus 2 disconnected from the rest of the network, and the remaining network different from the original network. Therefore, the prediction based on the synchronization condition may not be exact. The closest unstable equilibrium for DM is found, and the stability margin is found to be -34.87 (<0), which means the stability of the system is not certified. Similar to COM, the stability of the system is undetermined if a certificate is not issued.

Fig 8 (A) shows large deviations in the results for Generator 2 between the proposed analytic solution and TDS after $t$ = 1.5 s, but both indicate system instability. They both indicate that Generator 2 will lose synchronization quickly after the line fault, and be disconnected from the network (*Type IV, unstable*). In the on-fault condition, Generator 2 is isolated from the

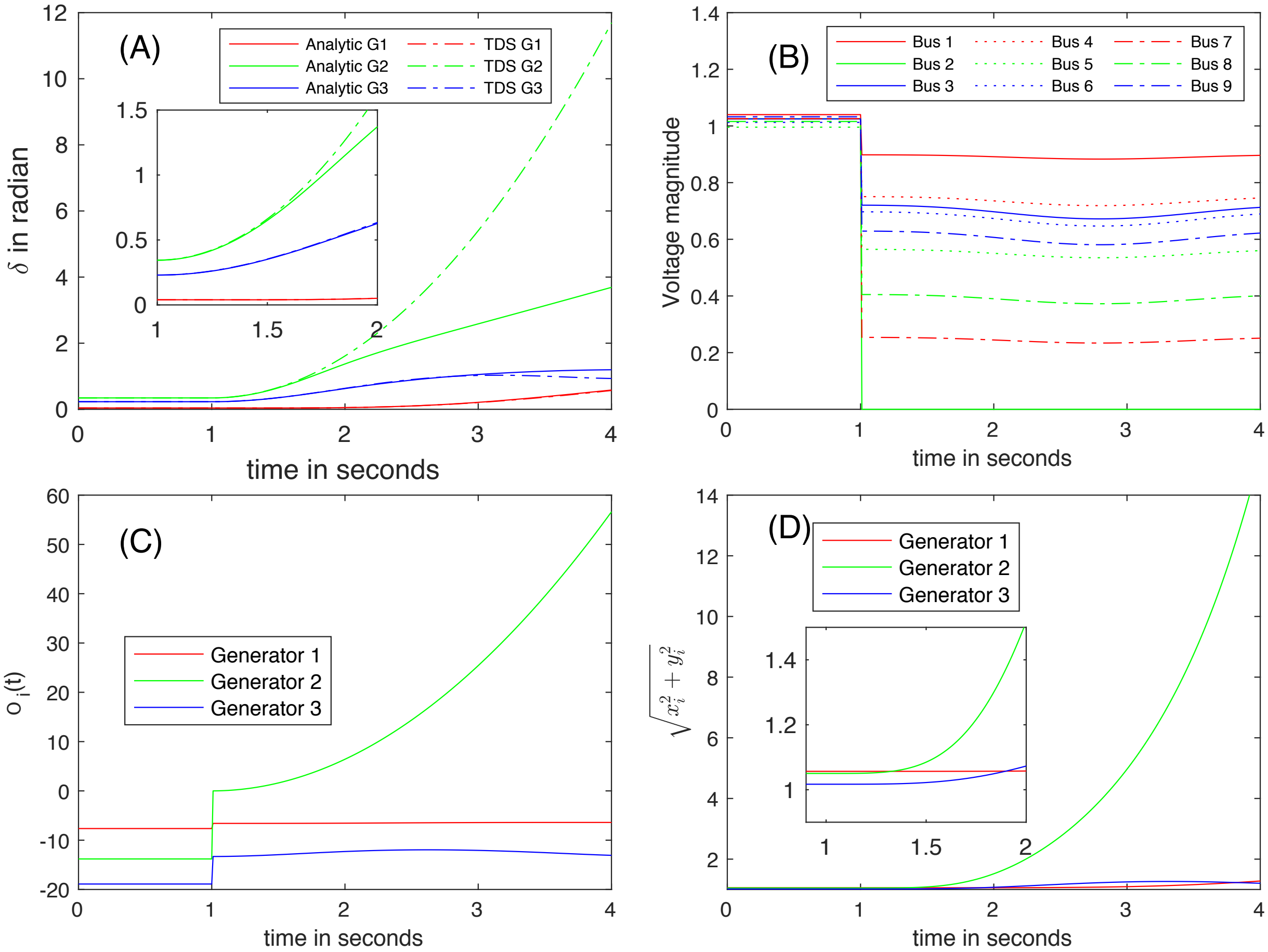

**Fig 8.** The trajectories of (A) $\delta$, of (B) $E$, of (C) $O_i(t)$, and of (D) the estimated magnitudes of internal voltages due to the line fault at the line between Buses 7 and 8 near Bus 7 at $t = 1$ sec (red cross in Fig 5).

rest of the system. Therefore, as shown in Fig 8 (A), the rotor angular motion becomes faster than motion of the other generators.

**Validity region.** Fig 8 (C) shows the fault-on trajectories of $O_i(t)$ if the fault is not cleared. $O_i(t)$ for Generator 2 is not a slowly varying function, and it does not follow the Karamata representation theorem because $\eta(t)$ divulges for Generator 2, meaning $\eta(t)$ is not bound. $\|T\|_F$ is 7.88, and the threshold for $\|\delta T\|_F$ is 7.88 × $10^{-2}$ (1% × 7.88), but the impact of O2 is 1.00, which is greater than its threshold. Fig 8 (D) lists the estimated magnitudes of the internal voltages. Shortly after the line fault, the internal voltages at Bus 2 increase too quickly, and O1 becomes large. Therefore, the validity region is limited quickly after 1 s, and neither O1 nor O2 are small during the on-fault trajectory.

## Line fault: post-fault trajectory

To prevent the loss of synchronization, the fault should be cleared. At $t = 1.1$ s, the fault is cleared by opening up the circuit breakers of the line between Buses 7 and 8. With the updated topology, the $Y_{bus}$ is updated; accordingly, a new operation point is identified. However, the rotor angular motions are continuous. Fig 9 show the post-fault trajectories of (A) rotor angle, and (B) estimated magnitudes of the internal voltages, respectively. Fig 9 (A) exhibits that the analytical approach does not predict the rotor dynamics properly after 2.5 s, and the discrepancy increases with time. The TDS has a time step of 0.01 s.

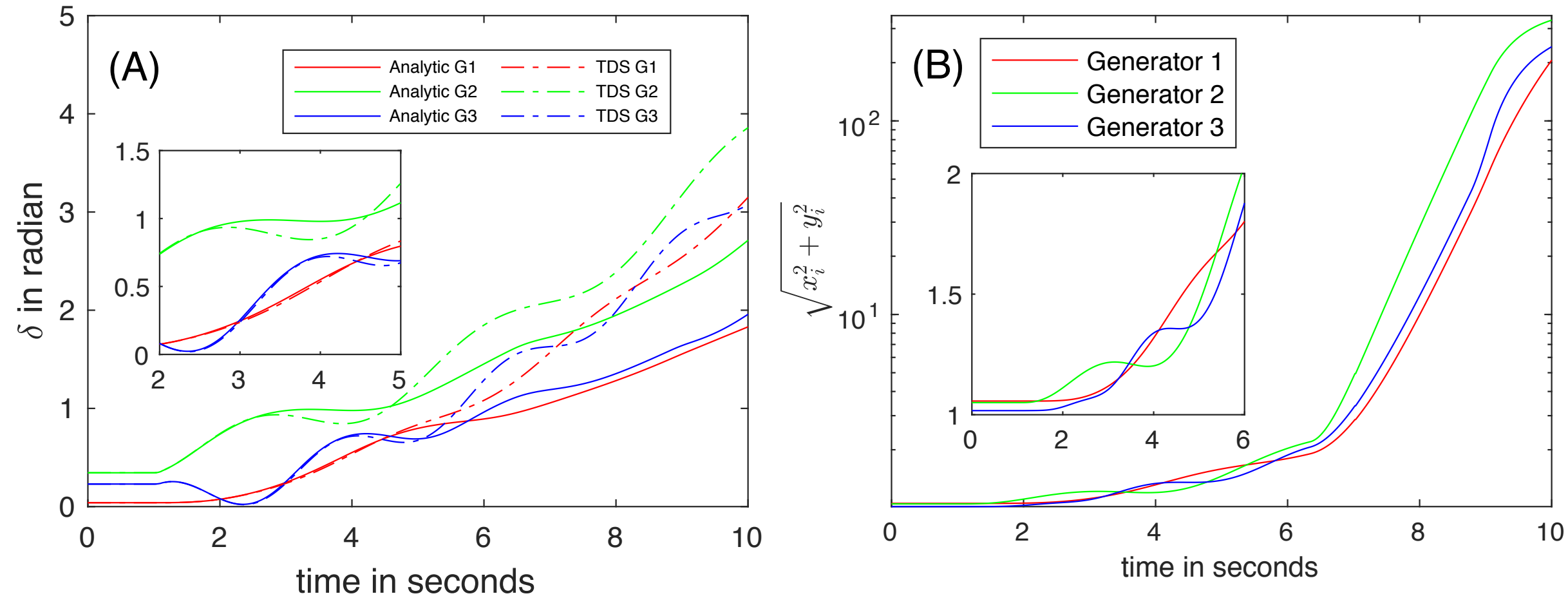

**Fig 9.** (A) Post-fault trajectories of the rotor angles and (B) the magnitudes of the estimated internal voltages.
https://doi.org/10.1371/journal.pone.0225097.g009

**Stability of the system.** The maximum angle difference in voltage angles immediately after clearing the fault is 0.155 rad, which COM concludes is the undetermined stability of the system. The closest unstable equilibrium point is observed, and the stability margin based on DM is -1.98 (<0), which also renders DM unable to estimate the stability of the system. There is a discrepancy in the results of the rotor angles between the proposed analytic method and TDS. This discrepancy increases after $t = 3$ s as the system evolves over time.

**Validity region.** The value of $\|T\|_F$ immediately after clearing the fault is 11.43; hence, the tolerance to the impact of O2 toward $\|\delta T\|_F$ is 0.114 (= 1% × 11.43). The impact of O2 is 1.79 ($\|\delta T\|_F / \|T\|_F = 0.156$), which is beyond the threshold of 1% after clearing the fault. Fig 9 (B) indicates that the voltage magnitudes suddenly increase after 2 s, meaning O1 is not small after 2 s. The validity region boundaries are identified with the projections, and the consistent initial points are identified to correct the errors.

## Post-fault trajectory: beyond the validity region

At the boundary of the validity region, $T$ and $z$ are modified according to (32) and (34). Because they are still close to the true ones at the boundary, the consistent initial points are evaluated at the boundaries. With the updated values for $T$ and $z$, one can compute the coefficient to construct the analytical solution under the constraints that O1 and O2 are small with the updated values. Fig 10 show the post-fault trajectories beyond the validity region of (A) router, (B) the magnitudes of the bus voltages, (C) $O_i(t)$, and (D) the estimated magnitudes of the internal voltages, respectively.

**Stability of the system.** Fig 10 (A) shows an improved similarity between two models. Fig 10 (B) illustrates the post-fault trajectories of the terminal voltages that indicate non-negligible changes of voltage magnitudes. The left and the right eigenvectors are identified corresponding to each eigenvalue $\lambda$ of $T$ to compute the condition of the eigenvalue, $s(\lambda)$ from (25). The largest deviation of the eigenvalue corresponds to the smallest condition of the eigenvalue:

$\max_j \left|\lambda_j - \hat{\lambda}_j\right| \approx \varepsilon \left[\min_j s(\lambda_j)\right]^{-1}$. It was found that the largest change in the eigenvalue was $\lambda_{\max}^{\mathrm{Re}} = 0.3879$ before the update, and the corresponding updated eigenvalue was $\hat{\lambda}_{\max}^{\mathrm{Re}} = 0.3734$, and $s(\lambda)$ was approximately 1.8. It turns out that the eigenvalue has the largest real part in the positive, but the corresponding coefficients are not large enough to make the system divulge before $t = 10$ s (*Type III* stable). As shown Figs 10 (A) and (B), both the analytical approach and TDS expect the stability of the system.

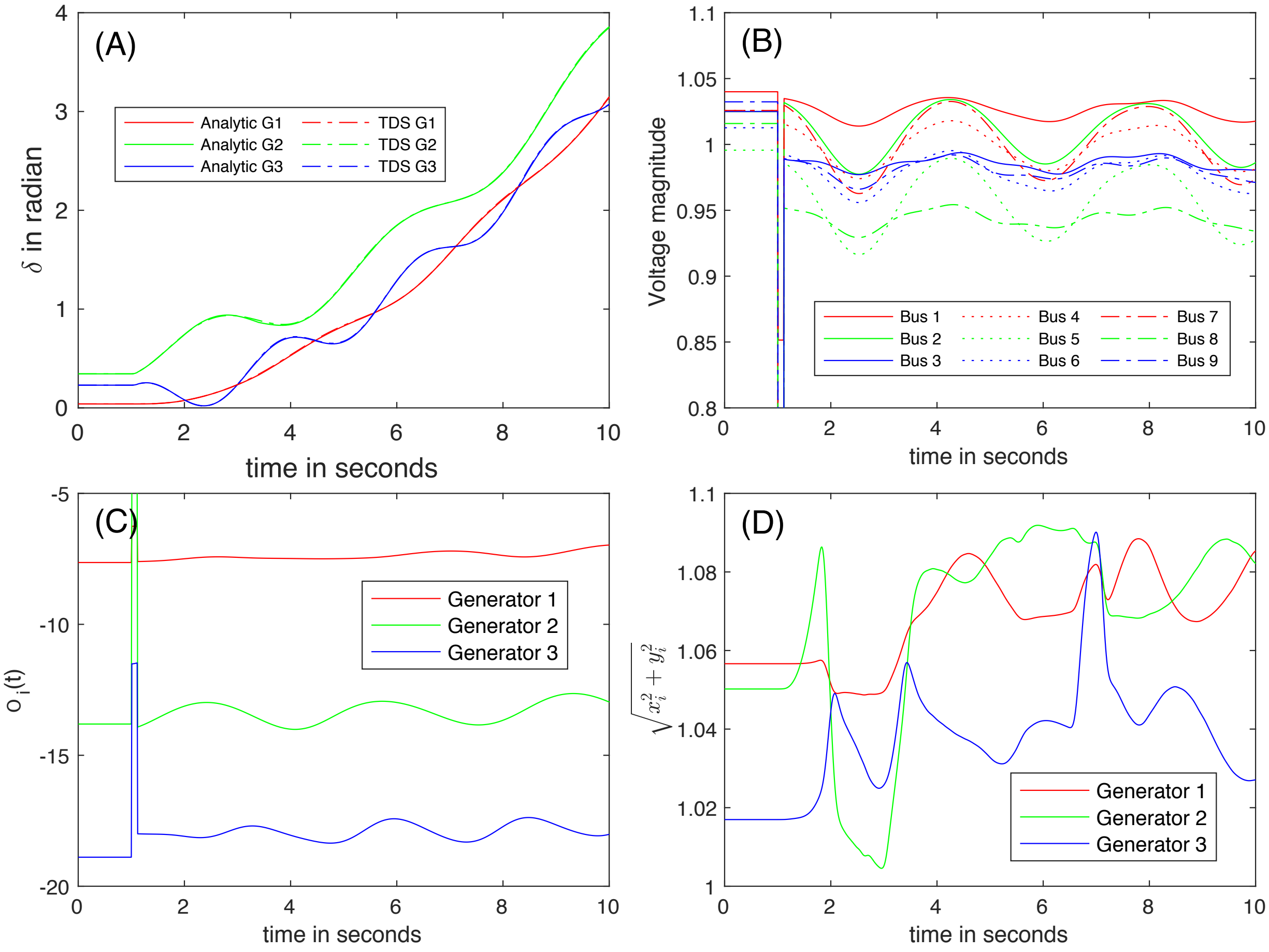

**Fig 10.** (A) Post-fault trajectories of the rotor angles and (B) trajectories of the voltage magnitudes, (C) $O_i(t)$, and (D) the estimated magnitudes of internal voltages after the fault is cleared at $t = 1.1$ seconds.
https://doi.org/10.1371/journal.pone.0225097.g010

**Validity region.** As discussed with Fig 10 (D), the validity region boundaries are frequently revisited (15 times), and consistent initializations are performed by the projections using (34). Note that the codes are not currently optimized to utilize the sparse structure of the matrices and to explain the partial update of the Jacobian matrices $\hat{M}_w$ and $\hat{M}_u$ in (34), which makes the computation process inefficient. The voltage magnitudes of Generator 2 swings widely because the generator is closest to the location of the line fault among all the generators.

## Future works

In this work, for the sake of simplicity, the dynamics of the generators are represented using the classical model. It is important to model the generators correctly to discuss the power system dynamics [23]. We plan to accommodate generator models that have been widely used for decades in numerous commercial simulation programs for modeling round-rotor and salient-pole synchronous generators [26] (the GENROU/GENSAL models) and for the detailed treatment of saturation (the GENTPF/GENTPJ models) [45]. The immediate adjustment in modeling is on *Ibus* where the voltage magnitude $E_i$ does not stay constant, which requires a modification in O1.

## Conclusions

In this paper, we derived an analytical solution to the swing equations to assess the transient stability of power grids. To the best of our knowledge, our solution is a unique analytical solution to the swing equations without physically unacceptable assumptions, while obeying Kirchhoff's laws. The solution indicates the factors affecting the stability after a disturbance occurs. Based on the solution, a new stability assessment approach is proposed. The assessment tool is different from the conventional assessment tools (COM, DM, and TDS approach) in that the derivation of our solution does not require the unphysical assumptions required for both COM and DM (such as lossless grids, constant voltages at all buses, and no consideration of reactive power). Moreover, its computational complexity is manageable. In addition to the low computational complexity, the approach proposed in this study explores the components affecting power system dynamics by examining the structure of the $T$ matrix in terms of system dependent ($T_{sys}$) and operation point dependent ($T_{op}$) submatrices. The simulation results show that O1 and O2 are small in most cases. However, even in a case when O1 and O2 are large, it is possible to maintain O1 and O2 small by introducing the validity region based on the projection methods. The consistent initializations make it possible to identify the trajectories reliably.

## Appendix

### Time domain simulation, TDS

For comparison, TDS is performed using the Runge–Kutta method [46]. At first, the loads are converted to equivalent admittance to construct a linear model (i.e., the current is linear with the internal voltages). From the linear current and voltage relationship, the parameters in (2) are evaluated at the post-fault state. Because the magnitudes of internal voltages are assumed constant, the voltage at the *Ibuses* can be evaluated in terms of δ. Once the voltages at the *Ibuses* are computed, one can compute all the terminal voltages and real power injection at the *Ibuses*. Using the terminal voltages and the real power injection, it is possible to update $\delta$.

We first define $z = \begin{bmatrix} \delta^T & \omega^T & \tilde{w}^T \end{bmatrix}^T$, and rewrite the swing equation: $y = \frac{dz}{dt} = Az + b$ where $A = \begin{bmatrix} 0 & I & 0 \\ 0 & -diag(D/M) & 0 \\ 0 & 0 & -\tilde{L} \end{bmatrix}, b = \begin{pmatrix} 0 \\ f(z) \\ -\tilde{l} \end{pmatrix}$, and $f(z) = diag(1/M)\left[p_{mech} - p_{elec}(z)\right]$. At the first sub step at iteration, with a step size $\Delta t > 0$, one finds

$$y_m^{\{1\}} = Az_m^{\{0\}} + b_{m-1}^{\{0\}}, \; z_m^{\{1\}} = z_{m-1} + \frac{1}{2}\Delta t y_m^{\{1\}} \quad \text{where} \quad z_m^{\{0\}} = z_{m-1}, \; b_{m-1}^{\{0\}} = b_{m-1} \tag{A1}$$

The update of $z$ allows finding a power flow solution corresponding to the value of $z$ for updating $f\left(z_m^{\{1\}}\right)$. (A1) is generalized at the $k^{\text{th}}$ sub step for the $m^{\text{th}}$ iteration as follows:

$$y_m^{\{k\}} = Az_m^{\{k-1\}} + b_{m-1}^{\{k-1\}}, \; z_m^{\{k\}} = z_{m-1} + r_k \Delta t y_m^{\{k\}} \quad \text{where} \quad r_1 = r_2 = \frac{1}{2}, r_3 = 1 \tag{A2}$$

The update of $b_{m-1}^{\{k\}}$ follows (A2). Then the last update at the $m^{th}$ iteration is

$$z_{m+1} = z_m + \frac{\Delta t}{6}\left(y_m^{\{1\}} + 2y_m^{\{2\}} + 2y_m^{\{3\}} + y_m^{\{4\}}\right) \tag{A3}$$

In averaging the four increments, a larger weight is given to the middle increments. Multiple simulations are performed with various time steps $\Delta t$ in a decreasing order until the simulation results are invariant.

### Direct method based on an energy function

The Lyapunov stability theorem gives sufficient conditions to determine the stability of the system. Numerical computation of the underlying ordinary differential equations is not necessary to derive the stability properties. The Lyapunov stability theorem gives two conditions for stability for a Lyapunov function $V$ on an open set $U$. (1) $V(x) = \begin{cases} = 0 & \text{if } x = 0 \\ > 0 & \text{otherwise} \end{cases}$ and (2) $\frac{dV(x)}{dt} = \sum_j \frac{\partial V}{\partial x_j}\frac{dx_j}{dt} = \nabla V \cdot \frac{dx}{dt} \le 0 \ \forall x \ne 0$.

For a power system with multiple machines, the traditional DM utilizes the energy function under the assumptions of losses ($\gamma_{ik} = 0$) and of the fixed voltage magnitudes ($E_k$ is fixed). The simplest energy function would be the mechanical interpretation [8] as follows:

$$V(\delta,\omega) = E_{KE} + E_{PE} = \frac{1}{2}\sum_k M_k\omega_k^2 + \left[-\sum_k p_k^{mech}\delta_k - \sum_{k,i}\left(|\tilde{y}_{ik}|E_iE_k\right)\cos\delta_{ik}\right] \tag{A4}$$

The corresponding swing equation is $\omega_i = \frac{d\delta_i}{dt}$, $M_i\frac{d\omega_i}{dt} = p_i^{mech} - \left(|\tilde{y}_{ik}|E_iE_k\right)\sin\delta_{ik} - D_i\omega_i$; therefore, $V$ satisfies stability conditions as follows:

$$\frac{dV(\delta,\omega)}{dt} = \sum_i\left[\left(\frac{\partial V}{\partial\delta_i}\right)\left(\frac{d\delta_i}{dt}\right) + \left(\frac{\partial V}{\partial\omega_i}\right)\left(\frac{d\omega_i}{dt}\right)\right] = -\sum_i D_i\omega_i^2 \le 0 \tag{A5}$$

The rotor angle $\delta^{ref}$ is the measure of the internal voltage angle with respect to the terminal voltage angle of the reference bus. For convenience, the reference frame for the rotor angle is redefined with respect to the rotating COI, i.e., $\delta_i = \delta_i^{ref} - \delta_{COI}$, $\omega_i = \omega_i^{ref} - \omega_{COI}$ where $\delta_{COI} = \frac{1}{M_T}\sum_i M_i\delta_i$, $\omega_{COI} = \frac{1}{M_T}\sum_i M_i\omega_i$, $M_T = \sum_i M_i$. Equilibrium points are found at a set of the rotor angles that satisfies $d\delta_i/dt = 0$ and $d\omega_i/dt = 0$, which leads to $\delta_{ik} = \sin^{-1}\left(\frac{p_i^{mech}}{|\tilde{y}_{ik}|E_iE_k}\right)$. Beside the stable equilibrium point, there will be the $NI$ neighboring unstable equilibrium points to the stable equilibrium point. At the unstable equilibrium points, the "kinetic energy" ($E_{KE}$) is zero, and the energy equals the "potential energy", $E_{PF}$. Along the unstable equilibrium points, one can find the closest unstable equilibrium point. At the point, the critical energy $V_{cr}$ is defined so that any trajectory starting from a point with a lower energy than $V_{cr}$ is guaranteed to converge into the stable equilibrium point if no other equilibrium points are contained in the set [10]. As a result, one can certify the system stability based on the stability margin, $V_{margin} = V_{cr} - V_{cl}$ where $Vcl$ is the current energy at the clearing time.

## Author Contributions

Conceptualization: HyungSeon Oh.

Data curation: HyungSeon Oh.

Formal analysis: HyungSeon Oh.

Investigation: HyungSeon Oh.

Methodology: HyungSeon Oh.

Software: HyungSeon Oh.

Validation: HyungSeon Oh.

Writing – original draft: HyungSeon Oh.

Writing – review & editing: HyungSeon Oh.